\documentclass[12pt,preprint]{emulateapj}

\usepackage{color}

\DeclareMathSymbol{\varOmega}{\mathord}{letters}{"0A}
\DeclareMathSymbol{\varSigma}{\mathord}{letters}{"06}
\DeclareMathSymbol{\varPsi}{\mathord}{letters}{"09}
\DeclareMathSymbol{\varPhi}{\mathord}{letters}{"08}

\newcommand{\Eq}[1]{equation (\ref{#1})}

\newcommand{\Fig}[1]{Fig.~\ref{#1}}

\shorttitle{Can Planetary Instability Explain the {\it Kepler} Dichotomy?}
\shortauthors{Johansen, Davies, Church, \& Holmelin}

\begin{document}

\title{Can Planetary Instability Explain the {\it Kepler} Dichotomy?}

\author{Anders Johansen, Melvyn B.\ Davies, Ross P.\ Church, and Viktor
Holmelin}
\affil{Lund Observatory, Department of Astronomy and Theoretical Physics,
Lund University, Box 43, 221 00 Lund, Sweden}

\begin{abstract}
The planet candidates discovered by the {\it Kepler} mission provide a rich
sample to constrain the architectures and relative inclinations of planetary
systems within approximately 0.5 AU of their host stars. We use the
triple-transit systems from the {\it Kepler} 16-months data as templates for
physical triple-planet systems and perform synthetic transit observations,
varying the internal inclination variation of the orbits. We find that all the
{\it Kepler} triple-transit and double-transit systems can be produced from the
triple-planet templates, given a low mutual inclination of around five
degrees. Our analysis shows that the {\it Kepler} data contains a population
of planets larger than four Earth radii in single-transit systems that can not
arise from the triple-planet templates. We explore the hypothesis that
high-mass counterparts of the triple-transit systems underwent dynamical
instability to produce a population of massive double-planet systems of
moderately high mutual inclination. We perform $N$-body simulations of
mass-boosted triple-planet systems and observe how the systems heat up and lose
planets by planet-planet collisions, and less frequently by ejections or
collisions with the star, yielding transits in agreement with the large planets
in the {\it Kepler} single-transit systems. The resulting population of massive
double-planet systems can nevertheless not explain the additional excess of
low-mass planets among the observed single-transit systems and the lack of
gas-giant planets in double-transit and triple-transit systems. Planetary
instability of systems of triple gas-giant planets can be behind part of the
dichotomy between systems hosting one or more small planets and those hosting a
single giant planet. The main part of the dichotomy, however, is more likely to
have arisen already during planet formation when the formation, migration or
scattering of a massive planet, triggered above a threshold metallicity,
suppressed the formation of other planets in sub-AU orbits.
\end{abstract}


\section{Introduction}

Radial velocity surveys searching for exoplanets have been operating for more
than 15 years since the first exoplanet discovery \citep{MayorQueloz1995}. One
of the most important discoveries of these surveys is that many stars host
multiple planets within a few AU
\citep[e.g.][]{Butler+etal1999,Wright+etal2009,Lovis+etal2011}. However, most
of the detected planetary companions are massive since such planets produce
higher radial velocity signals. Recent years have seen the discovery of a new
class of low-mass exoplanets -- super-Earths with masses less than ten times
the Earth's -- from radial velocity surveys
\citep[e.g.][]{Rivera+etal2005,Udry+etal2007}. Although the number of
super-Earths discovered by radial velocity surveys is growing rapidly
\cite[e.g.][]{Mayor+etal2011}, their number is not yet high enough to extract
statistical information about the abundance and general architectures of
planetary systems including low-mass planets.

The {\it Kepler} mission (which we will often refer to simply as {\it Kepler)}
provides the first statistically-significant survey of gas giants, Neptune-size
planets, super-Earths and terrestrial planets orbiting other stars
\citep{Borucki+etal2010}. Monitoring 156,000 stars for periodic dips in the
light curves, {\it Kepler} is sensitive to planets as small as one Earth radius
\citep{Fressin+etal2012}. The relative numbers of systems with single and
multiple transiting planets is a sensitive function of the intrinsic
multiplicity of planetary systems and the relative inclinations between planet
orbits and thus hold important information about planetary system architectures.

The target stars of {\it Kepler} are typically too faint for follow-up
observations of all but the largest planets, so detections by {\it Kepler} are
generally referred to as planet candidates, unless their masses can be
determined by radial velocity measurements \citep[e.g.][]{Batalha+etal2012} or
transit timing variations \citep[e.g.][]{Lissauer+etal2011a}. However, due to
the high photometric precision, fewer than 10\% of the planetary candidates are
believed to be false candidates \citep{MortonJohnson2011}. The strong
clustering of planet candidates into multiple systems is further evidence for
the physical nature of the detections, since false positives caused by binary
stars would be distributed evenly among the {\it Kepler} target stars
\citep{Lissauer+etal2012}.

The first four months of {\it Kepler} data revealed 1235 planetary candidates
\citep{Borucki+etal2011a} while that number had grown to 2321 after sixteen
months \citep{Batalha+etal2012} -- 253 of Earth size ($R \le 1.25 R_\oplus$),
712 of super-Earth size ($1.25 R_\oplus < R \le 2 R_\oplus$), 1078 of Neptune
size ($2 R_\oplus < R \le 6 R_\oplus$), 207 of Jupiter size ($6 R_\oplus < R
\le 15 R_\oplus$) and 71 with sizes larger than Jupiter ($15 R_\oplus < R <22
R_\oplus$). A large fraction of those planets are in multiple systems
\citep{Lissauer+etal2011b,Batalha+etal2012} -- 245 host stars show double
transits, 84 show triple transits, 27 show quadruple transits, 8 show quintuple
transits and 1 shows sextuple transits. \cite{OfirDreizler2012} used an
alternative data reduction algorithm to analyse the {\it Kepler} data and found
several new transit signals, upgrading one of the planet host stars to a new
candidate sextuple system.

The {\it Kepler} data has already led to new knowledge about the
characteristics of the innermost parts of planetary systems. Individual systems
such as the sextuple Kepler-11, with masses determined by transit timing
variations \citep{Lissauer+etal2011a}, show that small planets can have a
variety of densities and hence represent distinct classes of planets.
Statistical analysis of {\it Kepler} data gives constraints on formation and
migration processes. \cite{Youdin2011} showed that there is a difference
between the size distributions of shorter and longer period planets. The
apparent lack of planets of around three Earth radii in size, relative to both
smaller and larger planets, in orbits shorter than 7 days may be evidence of
either sublimation of volatiles or inefficient gas accretion in Neptune-mass
planets migrating to close orbits.

Transit observations are mainly biased by the orbital alignment with the line
of sight (affecting all planets) and stellar and instrumental noise (affecting
mainly smaller planets). The orbital alignment bias is relatively independent
of planet size, since stellar radius and planetary semi-major axis dominate the
transit probability. The transit bias is thus less complicated than the bias in
radial velocity detections. Together with the sensitivity down to Earth-sized
planets, this makes the {\it Kepler} data well suited for studies of planetary
system architectures.

The goal of this paper is to find underlying planetary system architectures
that explain the absolute and relative number of single and multiple transits
in the {\it Kepler} data. Our approach differs from that of other authors.
\cite{Lissauer+etal2011b} used the {\it Kepler} data to parameterise the
population of planets in terms of their sizes and semi-major axes. They found
that the majority of the observed double-transit systems and a substantial
fraction of the observed single-transit systems could arise from relatively
flat systems of higher multiplicity, while the remaining single-transit systems
come from a second distinct population. \cite{TremaineDong2011} looked at the
problem from a statistical approach. They concluded that the inversion from
observed transits to underlying populations is in principle degenerate in the
mutual inclination parameter, in that solutions consisting of a combination of
single-planet systems and very densely populated systems (up to 40 planets
within 0.5 AU) of nearly isotropic orbits can yield the observed transit
rates. It is nevertheless not clear that these nearly isotropic solutions are
physical, in that most planets need to be in very densely packed systems, with
few or no planetary systems of low multiplicity. \cite{Weissbein+etal2012}
used an approach similar to \cite{Lissauer+etal2011b} and find that their
assumption that planetary occurrence is an independent statistical process is
not supported by the data. In this paper we use an alternative approach to
constraining underlying planetary system architectures. We use the
triple-transit systems observed by the {\it Kepler} mission as templates for
physical planetary systems and perform synthetic transit observations to derive
the relative number of single, double and triple transits as a function of the
mutual inclination between the planet orbits. We thus work close to the
original data and do not have to go through steps of interpretation and
analysis of the statistical properties of planet sizes and orbits to fit these
with distribution functions.

Our paper is organized as follows. In \S\ref{s:selection} we explain how we
select the underlying planet population by using the triple-transit systems
observed by the {\it Kepler} mission as templates. In \S\ref{s:rates} we
perform synthetic transit observations of these systems, weighting each
template system by the inverse of the triple-transit probability. We compare
the synthetic transits with the {\it Kepler} data in \S\ref{s:fit} and find
that there is a good match in radii and semi-major axes between the synthetic
double and triple transits and the {\it Kepler} data. However, the synthetic
single transits do not match those observed -- the synthetic catalogue contains
too few single transits (by about a factor of three) and it completely lacks
planets having large radii (between the sizes of Neptune and Jupiter) which
have been found by the {\it Kepler} mission. This suggests that in addition to
the population of three-planet systems (seen in double and triple transits),
there is a second, distinct population of planetary systems containing larger
planets. We discuss the properties of this second population in
\S\ref{s:second}. We explore the hypothesis that planetary systems form with a
range of masses and that the most massive are inherently unstable to
planet-planet interactions. We perform $N$-body simulations of mass-boosted
versions of the {\it Kepler} triple-planet templates in \S\ref{s:stability}.
The typical result is that two planets collide and merge and leave a moderately
inclined double-planet system. While this second population in principle can
produce the large planets seen in single-transit systems in the {\it Kepler}
data, this does not explain an additional excess of small planets (smaller than
four Earth radii) among the single-transit systems, nor the lack of gas giants
in double-transit and triple-transit systems. In \S\ref{s:multiple} we relax
the assumption that all planetary systems are triple, constructing
single-planet and double-planet systems by removing one or two planets from the
triple-planet templates, and show that this does not change the necessity for a
dichotomy of planetary systems in nature to explain the data. We conclude in
\S\ref{s:conclusions}, proposing that the main part of the dichotomy between
systems showing single transits and systems showing double transits and triple
transits arose already during planet formation, when the migration or formation
of a large planet suppressed the formation of additional planets in sub-AU
orbits.

\section{Selection of Underlying Planet Population}
\label{s:selection}

We will in this paper make the assumption that an $i$-multiple system observed
by the {\it Kepler} mission can be used as a template for a physical
$i$-multiple system. A system with $i$ transiting planets observed by the {\it
Kepler} mission may in reality have more planets than the observed $i$,
especially further from the star where the transit probability is low, but a
reduction in planet number produces a system that is equally physical, as such
reduction very rarely leads to dynamical instabilities. We sometimes use the
notation $i$t-$j$p for observing $i$ transiting planets in an intrinsically
$j$-multiple system. 

To get a uniform and unbiased sample of planetary systems, we first apply a
number of selection criteria to the planets in the 16-months {\it Kepler} data:
\begin{enumerate}
  \item The signal-to-noise ratio must be larger than 16.
  \item The planet period must be shorter than 240 days.
  \item The planet radius must be smaller than 22.4 Earth radii.
\end{enumerate}
The first criterion ensures that the sample is close to being complete and that
no size bins are dominated by fortuitous detections, an important issue when
comparing observations of synthetic planetary systems to the {\it Kepler} data.
We follow \cite{Lissauer+etal2011b} and choose a limiting signal-to-noise ratio
of 16. The second criterion includes only planets seen to transit at least
twice in the 16-month data. The third criterion ensures that only
planetary-mass objects are selected, applying a maximum radius of $22.4$ Earth
radii, as proposed by \cite{Borucki+etal2011b}.

Selection criteria are applied on a planet-by-planet basis, so dense systems
may become sparser after the selection criteria. The triple system KOI-284 has
two planets in very similar orbits. We exclude this system because the two
planets likely orbit the two different stellar components of a binary
\citep{Howell+etal2011,Lissauer+etal2012}. Another peculiar system is KOI-191,
an originally quadruple system with a Jupiter-sized and an Earth-sized planet
very close to each other. However, the small planet has a low signal-to-noise
ratio and is excluded by criterion 1 above, reducing KOI-191 to a regular
triple-transit system. \cite{Fabrycky+etal2012} reported a similarly curious
quadruple system, KOI-2248, with two very close planets, but the detections of
two of the four planets have a low signal-to-noise ratio, and hence KOI-2248
reduces to a regular double-transit system after applying our selection
criteria.

\begin{figure*}
  \begin{center}
    \includegraphics[width=0.8\linewidth]{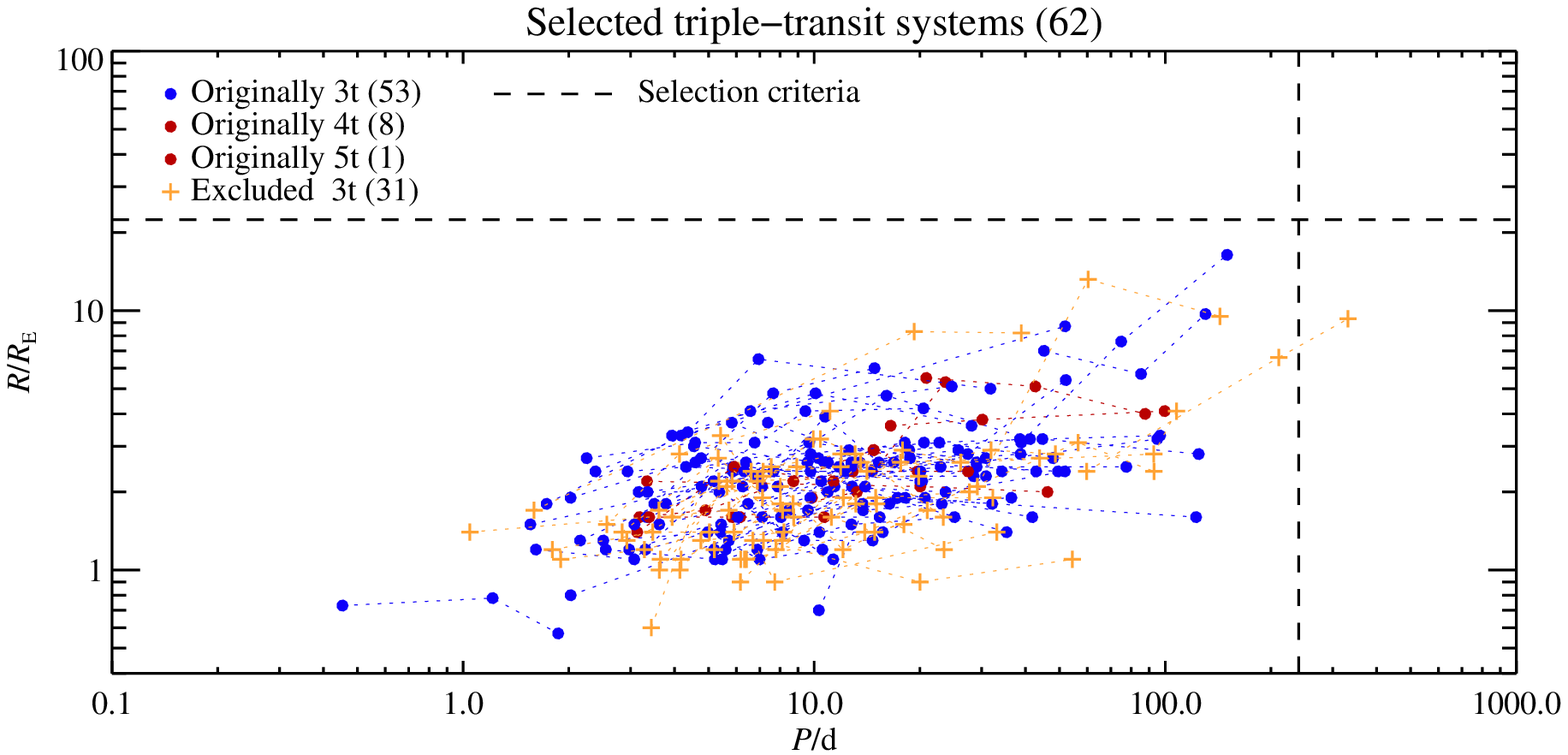} \\
    \includegraphics[width=0.8\linewidth]{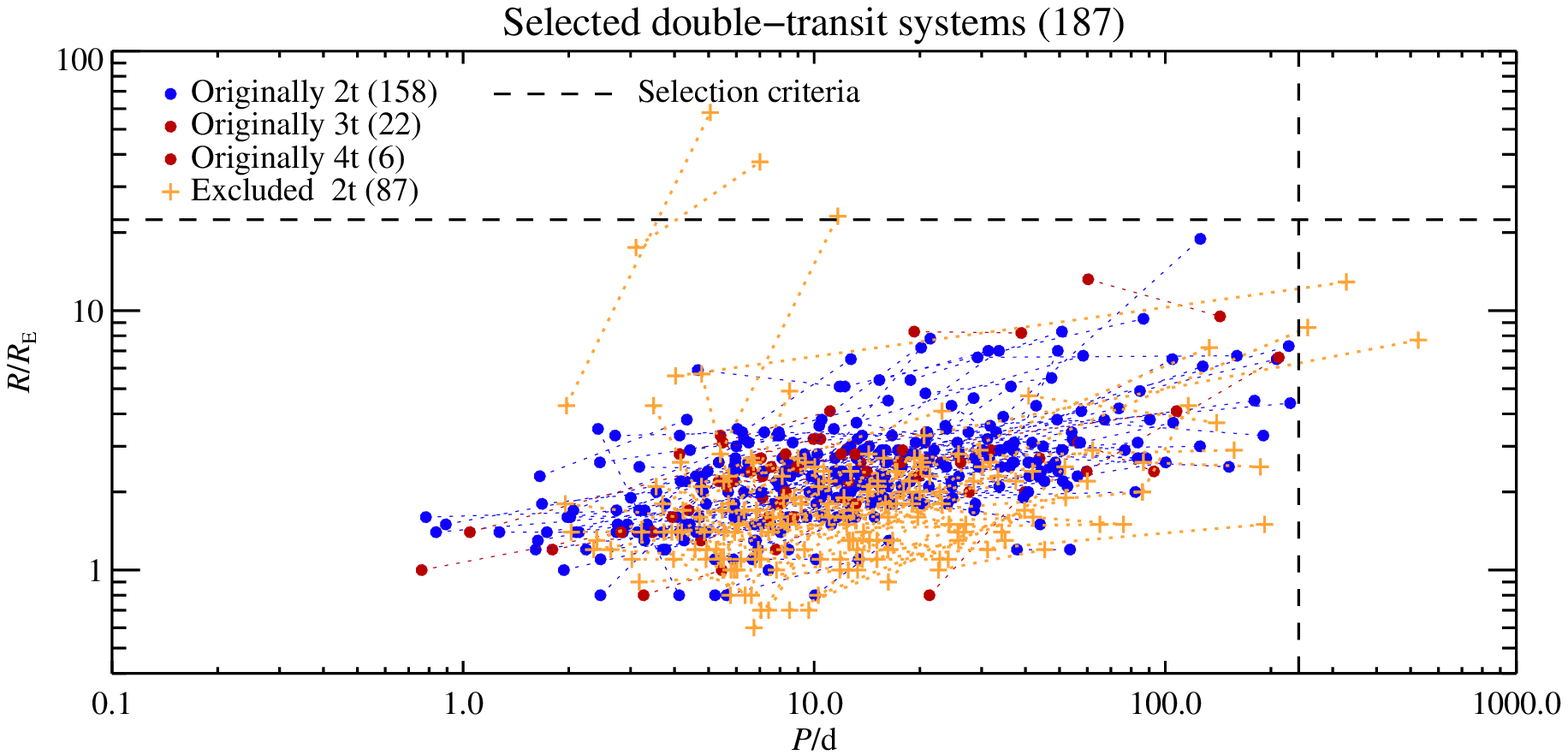} \\
    \includegraphics[width=0.8\linewidth]{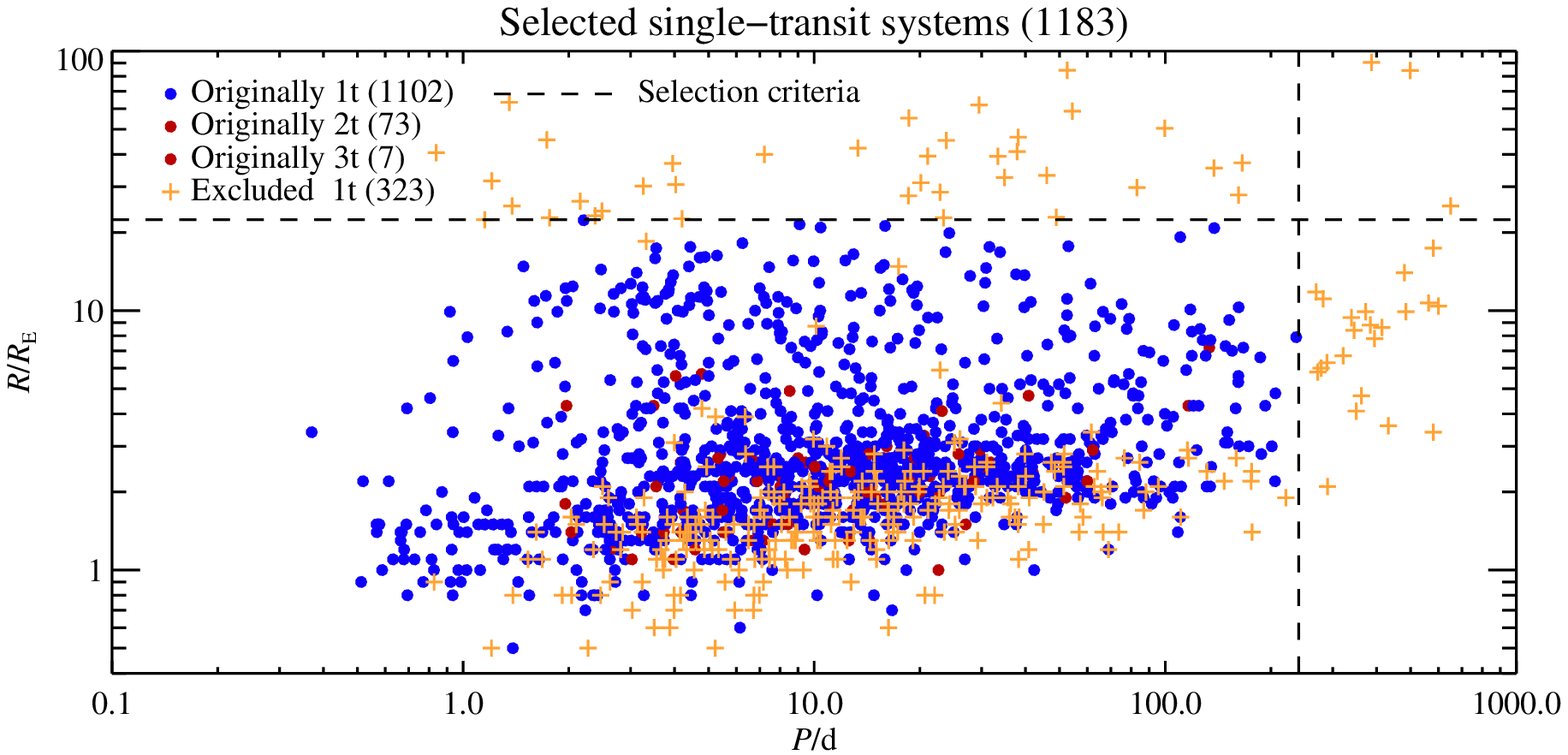}
  \end{center}
  \caption{Triple-transit, double-transit and single-transit systems in the
  16-months {\it Kepler} data after applying our selection criteria. Selection
  criteria for maximum orbital period and planet radius are indicated with
  dashed lines. The third selection criterion for signal-to-noise ratio
  generally excludes small planets. Systems reduced in planet number are marked
  with a yellow plus, while filled circles indicate systems that had originally
  the same number of planets (blue circles) or more planets (red circles).}
  \label{f:selected}
\end{figure*}

\Fig{f:selected} shows the triple-transit, double-transit, and single-transit
systems present in the {\it Kepler} data after applying selection criteria. We
do not consider higher-order systems since these are too few to serve as a
statistically significant template base for the synthetic systems. The number
of triple systems is reduced from originally 84 to 62, the number of double
systems is reduced from 245 to 187, while the number of single systems is
reduced from 1425 to 1183.

\begin{figure*}
  \begin{center}
    \includegraphics[width=\linewidth]{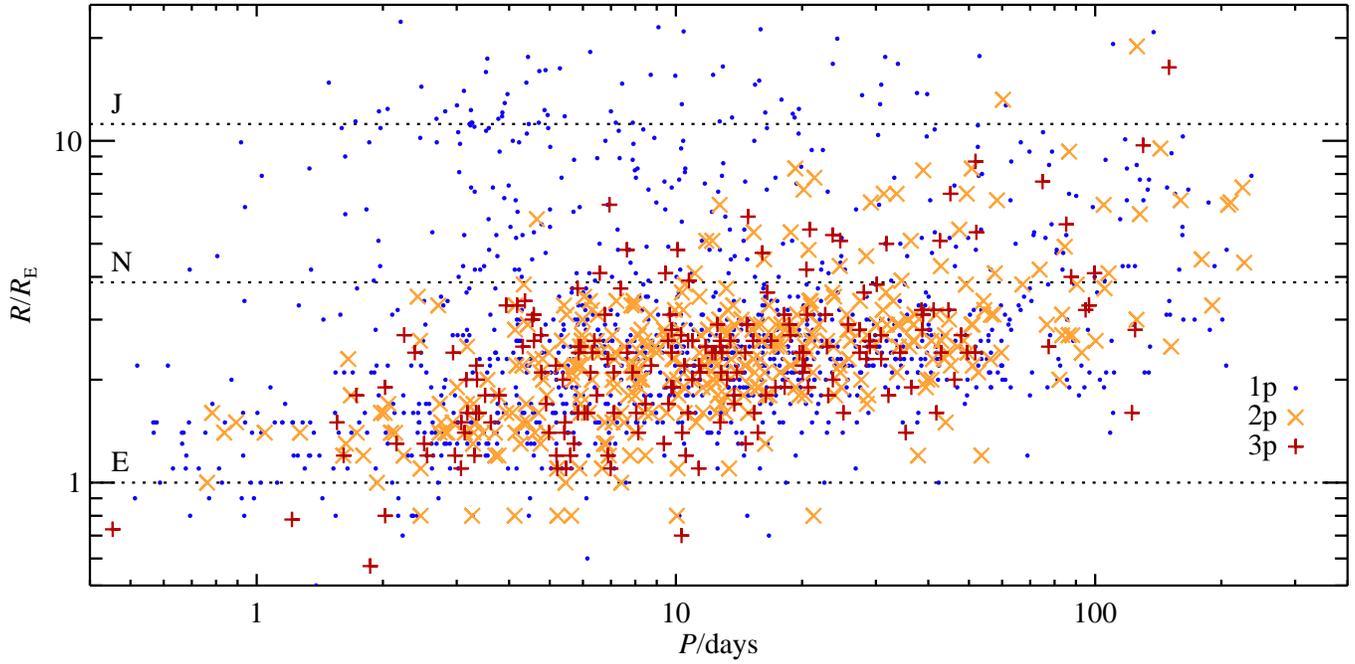}
  \end{center}
  \caption{Planetary radius versus orbital period of the selected
  triple-transit, double-transit and single-transit systems. Planets in
  double-transit and triple-transit systems are statistically similar, while
  planets in single-transit systems are on the average significantly larger
  than in more crowded systems.}
  \label{f:Rp_P}
\end{figure*}

The orbital periods and radii of the selected planets are shown in
\Fig{f:Rp_P}. While planets in systems with two or three transits are
statistically similar, planets in systems with only a single transit are
clearly on average significantly larger than the planets in higher-order
systems \citep{Latham+etal2011,Lissauer+etal2011b}.

\section{Calculation of Transit Rates}
\label{s:rates}

The goal of the paper is to perform simulated transit observations of
populations of synthetic planetary systems to find the intrinsic population of
planetary systems that matches the {\it Kepler} data. We use the triple
planetary systems observed by {\it Kepler}, after application of selection
criteria to make the sample uniform (see \S\ref{s:selection}), as templates for
physical systems and assume that stars have either zero planets or a planetary
system with the radii and semi-major axes of one of the triple-planet systems
observed by {\it Kepler}. Although crude, the assumption of planetary systems
having either zero or three planets turns out to give a very good fit to the
observed double-transit systems. Our approach is also very straight-forward to
apply to the data, in that we do not need to go through complex steps of
extracting statistical information about planetary sizes and semi-major axes
from the data. We will later relax the assumption that planetary systems are
always triple and show that this does not change the conclusions (see
\S\ref{s:multiple}).

All the template planetary ``systems'' are limited to a maximum semi-major axis
of approximately 0.5 AU. While this is much smaller than the expected total
extent of planetary systems in general, the {\it Kepler} data shows a
surprising population density and variety in these relatively close orbits. We
choose triple-transit systems as templates, since denser systems (quadruple and
beyond) are observed in very low numbers and are hence much more affected by
small number statistics than the triple-transit systems. As discussed in the
introduction, \cite{TremaineDong2011} showed that very densely populated
systems of nearly isotropic orbits can in principle explain the {\it Kepler}
data, but we will in this paper assume that systems more dense than three
planets are rare and hence do not contribute significantly to the triple
transits in the {\it Kepler} data. This assumption is discussed further in
\S\ref{s:conclusions}.

The orbits of all three planets are initially put in the same plane. We assume
that the orbits have very low eccentricities between 0 and 0.01 \citep[a good
approximation to the statistical properties of the {\it Kepler} transit
durations, see][]{Moorhead+etal2011}. For each planet in a triple system we
then choose an inclination angle distributed evenly between 0 and $\beta$. Here
$\beta$ is a measure of the opening angle of the system. We incline the orbital
plane of the planet by this angle relative to the initial, common orbital
plane. Afterwards we choose a random angle between 0 and $2 \pi$ and rotate
around the axis perpendicular to the original plane to get a random longitude
of ascending node and argument of periapsis.

\begin{figure}
  \begin{center}
    \includegraphics[width=\linewidth]{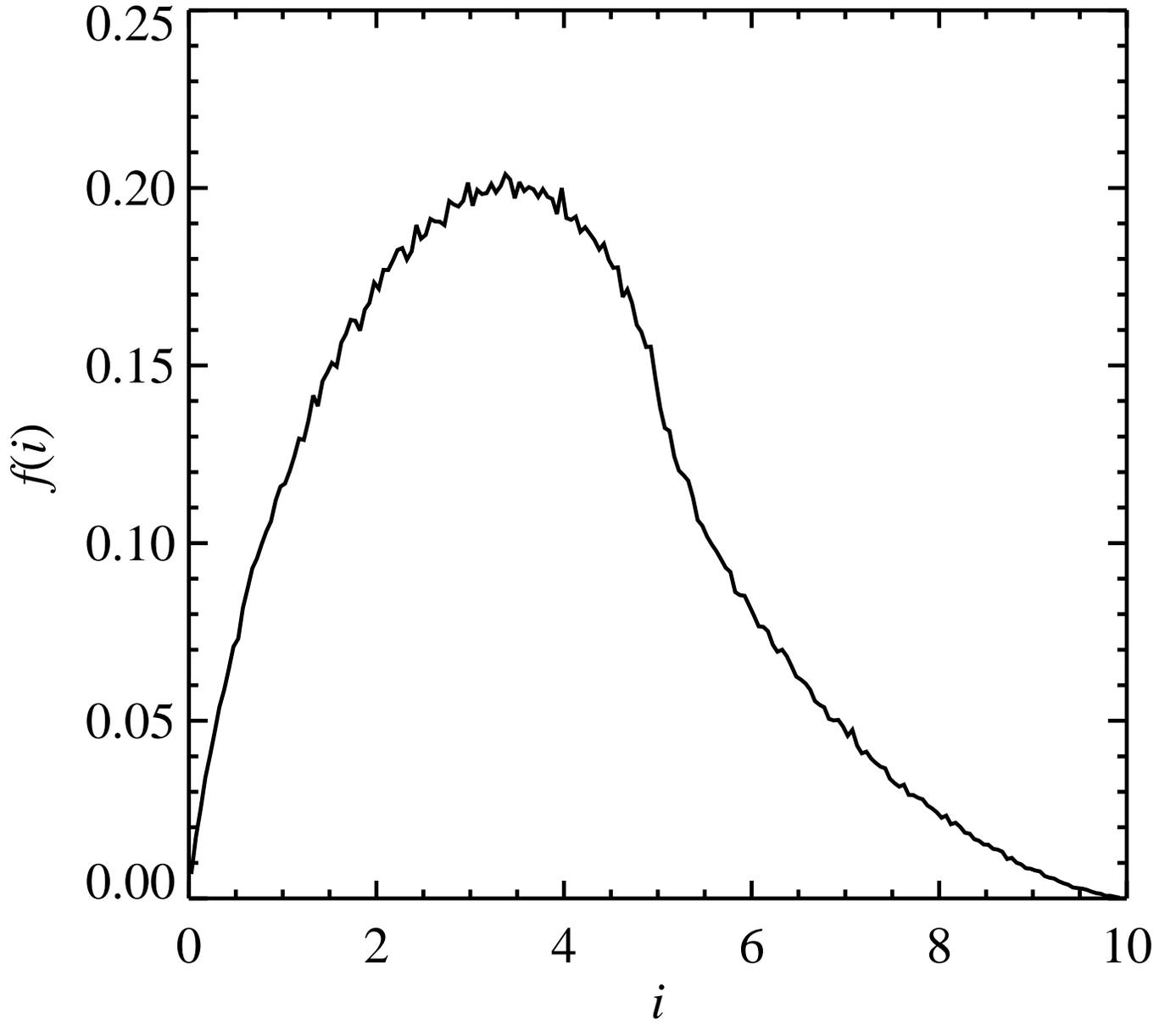}
  \end{center}
  \caption{The probability distribution function of the mutual inclination
  between pairs of planetary orbits inclined by a random angle distributed
  evenly between 0 and $\beta=5^\circ$ and with a random longitude of ascending
  node. The absolute maximum relative inclination is $i=2\beta$, obtained for a
  planet pair whose orbits are both inclined by the angle $\beta$ but with
  opposite longitude of ascending node. The distribution peaks at a mutual
  inclination around $i\approx3.5^\circ$.}
  \label{f:inclination_distribution}
\end{figure}

\Fig{f:inclination_distribution} shows how a chosen $\beta=5^\circ$ yields a
distribution in relative inclinations between planet pairs. We have taken pairs
of planetary orbits and inclined and rotated the planes as described above. The
distribution peaks at a mutual inclination around $i\approx3.5^\circ$. The
maximum relative inclination between any two planets, $2 \beta$, is obtained by
first inclining both planetary orbital planes by $\beta$ and then choosing
opposite longitudes of the ascending node.

We calculate the probabilities for an observer to see each of the planets
transiting by considering 1,000 random realisations of each triple-planet
system, for a given value of $\beta$, from a large number of random
directions. We choose directions uniformly distributed on the unit sphere and
calculate the total number of single, double and triple transits as well
identifying which planets are involved.

Using $i$t systems as templates for $i$p systems requires us to take into
account that the $i$t system was observed at all. Take for example an observed
system with three planets close to the star and another observed system with
three planets far away from the star. These two transiting systems can be used
as templates for two physical systems. However, if the wide system is ten times
less likely to be observed than the tight system, then we must convert the two
observed transiting systems into one tight synthetic planetary system and ten
wide synthetic planetary systems.

We need to go through a number of steps from the transit probability of
individual planets to calculating the number of single, double and triple
transits that are observed. We use a notation defined as following.
\begin{itemize}
  \item Probability that planet $i$ transits in system $j$:
\end{itemize}
\begin{equation}
  p_{ij} = \frac{R_{\star j}}{a_{ij}} \label{eq:ptransit} \, .
\end{equation}
\begin{itemize}
  \item Probability of exactly {\it i} transits in system {\it j}:
\end{itemize}
\begin{equation}
  P_{ij} = P_{ij}(\beta) \, . \label{eq:ptransit2}
\end{equation}
\begin{itemize}
  \item Mean weighted probability of an $i$-transit:
\end{itemize}
\begin{equation}
  P_i = \frac{\sum_j P_{ij} W_j}{\sum_j W_j} \label{eq:pweighted} \, .
\end{equation}
\begin{itemize}
  \item Weight of system $j$:
\end{itemize}
\begin{equation}
  W_j = \frac{1}{P_{3j}} \label{eq:weight} \, .
\end{equation}
Here $R_{\star j}$ is the radius of the host star in system $j$ and $a_{ij}$ is
the semi-major axis of planet $i$ in system $j$, with $i=1$ denoting the inner
planet, $i=2$ the middle, and $i=3$ the outer planet. The probability $p_{ij}$
that planet $i$ is seen to transit in system $j$ is then given by
\Eq{eq:ptransit}, ignoring the small contribution of the planetary radius to
the transit probability. The probability that system $j$ shows exactly $i$
planets transiting is denoted $P_{ij}$ (equation \ref{eq:ptransit2}). This
probability is a (complicated) function of the mutual inclination parameter
$\beta$. So far we have not considered the relative weight of the systems, but
as explained above, the fact that the {\it Kepler} mission observes a given
triple-transit system automatically requires that systems with low
triple-transit probabilities are more frequent among the target stars than
systems with high triple-transit probabilities. Hence we define $P_i$ as the
mean weighted probability that any given host star shows exactly $i$ planets
transiting, given by a straight-forward weighting in \Eq{eq:pweighted}. The
weighting function is defined in \Eq{eq:weight}. This choice ensures that each
triple-transit system in the {\it Kepler} data will be observed exactly once in
the synthetic observations. Effectively $W_j=1/P_{3j}$ is the number of
randomly oriented copies of system $j$ among the {\it Kepler} target stars. We
thus implicitly assume that higher-order planetary systems do not contribute
significantly to the observed triple-transit systems.

\begin{figure*}
  \begin{center}
    \includegraphics[width=0.8\linewidth]{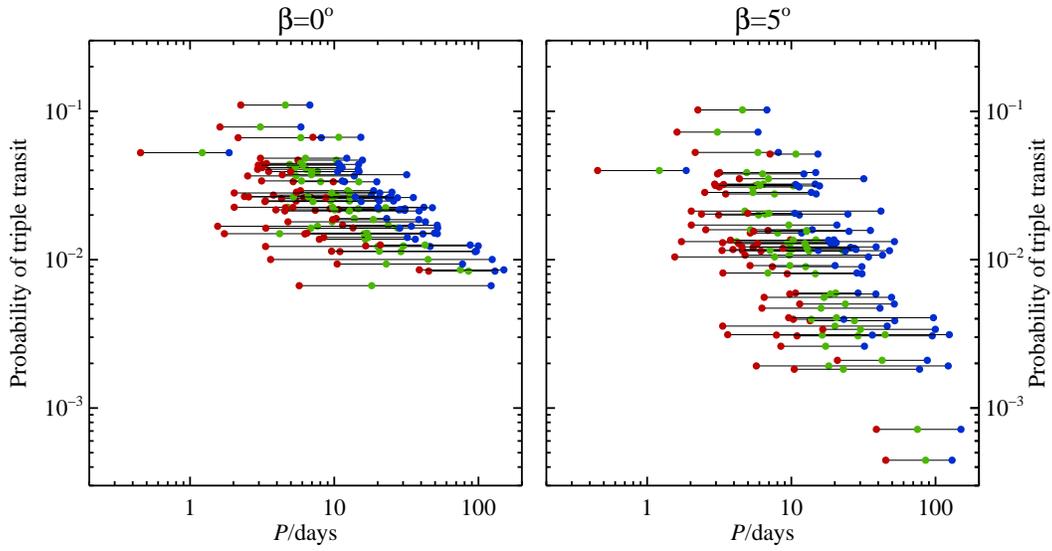}
  \end{center}
  \caption{The probability for the selected triple-planet systems to be seen in
  triple transit, for two values of the mutual inclination parameter $\beta$.
  The system with the highest triple-transit probability, approximately
  11\% at $\beta=0^\circ$, is KOI-1835, a relatively large star orbited by
  planets in 2.2, 4.6 and 6.8 day orbits. The system with the lowest transit
  probability has a more than ten times lower probability for $\beta=0^\circ$
  and would hence be 10 times more frequent among the synthetic systems. The
  probability contrast increases by more than a factor ten when increasing the
  mutual inclination parameter to $\beta=5^\circ$, as planets in wide orbits
  are much more likely to be inclined out of view than planets in tight orbits.}
  \label{f:weighting}
\end{figure*}

The probabilities of observing each of the {\it Kepler} triple-transit systems,
$P_{3j}$ using the above notation, are shown in \Fig{f:weighting}. The highest
triple-transit probability -- approximately $11.2\%$ at $\beta=0^\circ$ and
$10.1\%$ at $\beta=5^\circ$, is obtained for the KOI-1835 system, with three
close-in planets orbiting a relatively large host star of radius 1.66
$R_\odot$. Another noteworthy system is KOI-961 which has three small planets
very close to the star, with periods 0.45 days, 1.22 days and 1.87 days.
However, the probability of {\it Kepler} seeing three transits from this system
is not as high as one might expect, since the radius of the host star is low
\citep[0.17 solar radii, see][]{Muirhead+etal2012}. The contrast between the
lowest and highest triple-transit probability is around 10 for a mutual
inclination parameter of $\beta=0^\circ$ and 250 for $\beta=5^\circ$. Planets
in wide orbits are much more likely to be inclined out of view as $\beta$ is
increased, compared to planets in tight orbits.

\section{Fitting the Transits of the {\it Kepler} Mission} \label{s:fit}

\begin{figure}
  \begin{center}
    \includegraphics[width=\linewidth]{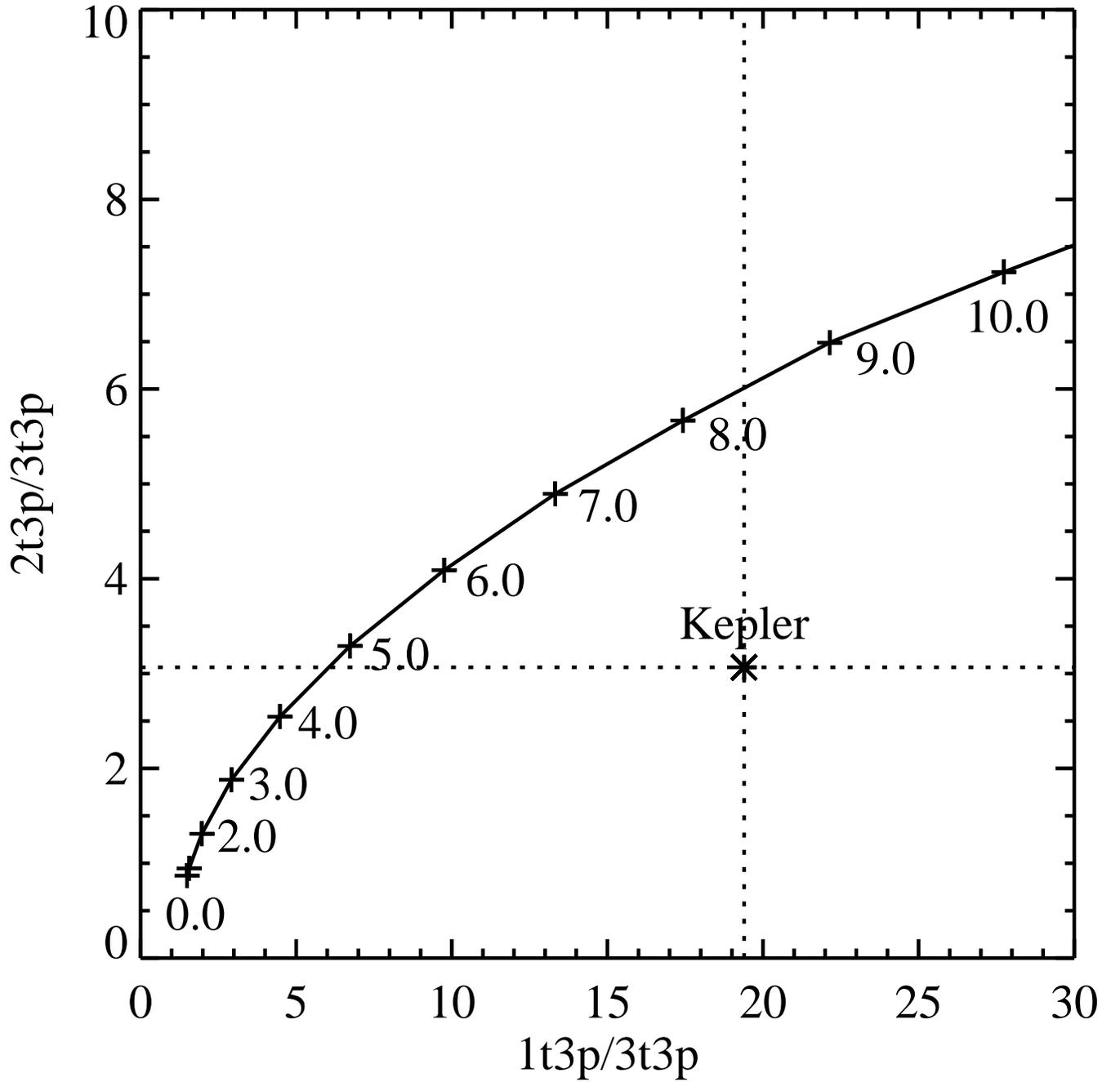}
  \end{center}
  \caption{The number of double-transit systems from the synthetic planet
  population versus the number of single-transit systems, divided by the number
  of triple-transit systems, for different values of the mutual inclination
  parameter $\beta$. The same ratio in the {\it Kepler} data, after applying
  selection criteria, is indicated by a star. It is not possible for any value
  of $\beta$ to get simultaneously the observed fraction of double-transit and
  single-transit systems.}
  \label{f:n2n3_n1n3}
\end{figure}

The transit detection algorithm described in \S\ref{s:rates} allows us to
calculate the number of single, double and triple transits arising from the
triple-planet templates and to compare the results with the {\it Kepler} data.
We vary the mutual inclination parameter $\beta$ between 0 and 10 degrees, in
steps of 1 degree. Each synthetic triple-planet system is given a weight equal
to the inverse triple-transit probability at the given inclination parameter.
The resulting numbers of single-transit and double-transit systems produced are
shown in \Fig{f:n2n3_n1n3}. We normalize the number of single-transit and
double-transit systems by the number of triple-transit systems. Matching those
ratios to the {\it Kepler} data, the correct absolute number of transits can be
obtained by assuming that a fraction $f=(N_3/P_3)/N_\star$ of the $N_\star$
{\it Kepler} target stars host triple-planet systems. It is clear from
\Fig{f:n2n3_n1n3} that we can not simultaneously reproduce both the number of
double-transit and single-transit systems in the {\it Kepler} data. Reproducing
the correct fraction of single-transit systems overproduces the number of
double-transit systems by a factor two, while a faithful reproduction of
double-transit systems will only produce 1/3 of the observed single-transit
systems. An important property of fitting synthetic transits to observed
transits is that no system produces a negative number of transits. Hence no
population of planets is allowed to overproduce the number of single, double or
triple transits found in the {\it Kepler} data. That puts a hard limit of
$\beta \lesssim 5^\circ$ to avoid overproduction of double-transit
systems\footnote{This limit relies on our assumption that planetary systems
with intrinsic multiplicity larger than three are rare. Considering templates
of higher multiplicity would lead to a higher value of $\beta$
\citep{TremaineDong2011}. However, this would be in some conflict with the
measured coplanarity of systems such as the sextuple Kepler-11 where the mutual
inclination is around one degree \citep{Lissauer+etal2011a}.}.

\begin{figure*}
  \begin{center}
    \includegraphics[width=\linewidth]{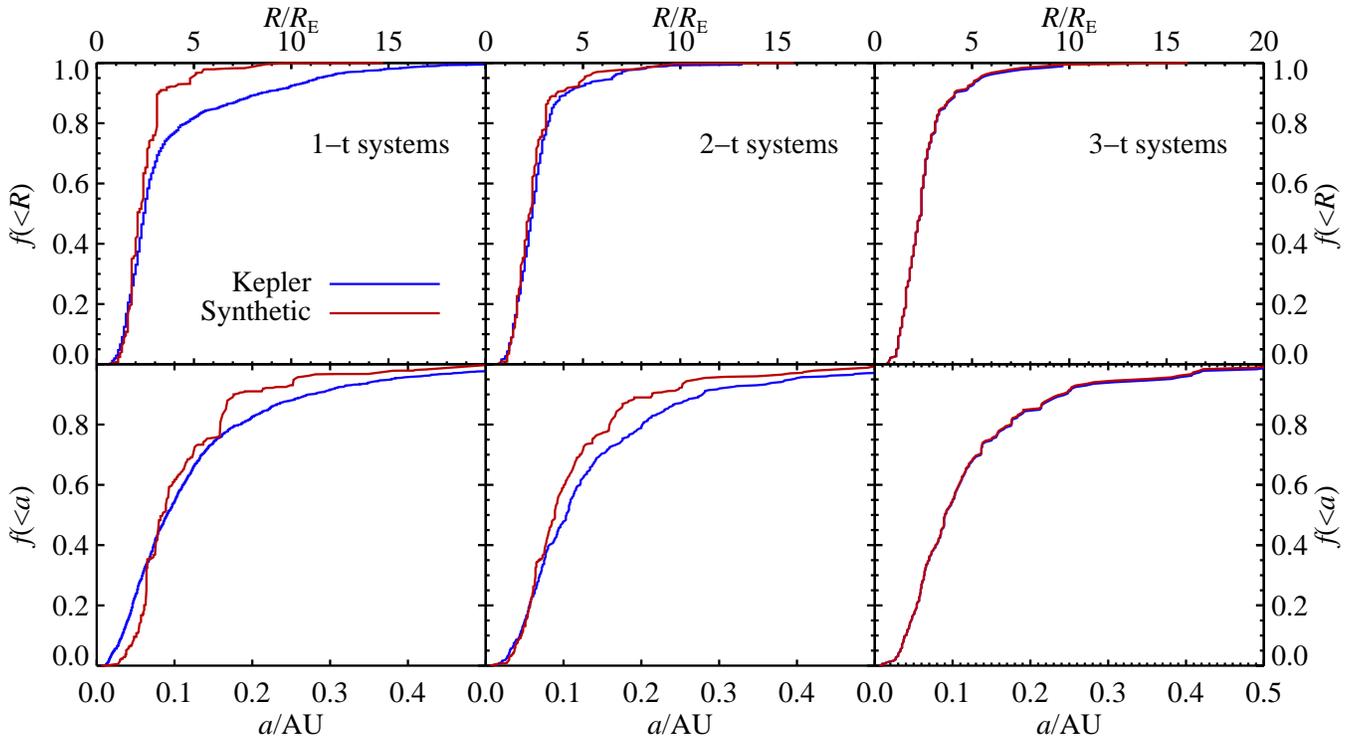}
  \end{center}
  \caption{Cumulative distribution functions of planetary radii (top) and
  semi-major axes (bottom) of synthetic transits (red) and of {\it Kepler}
  transits (blue) in single-transit systems (left panel), double-transit
  systems (middle panel) and triple-transit systems (right panel). We have
  chosen a mutual inclination parameter $\beta=5^\circ$ that approximately
  matches the ratio of double to triple transits in the {\it Kepler} data. The
  synthetic transits match the radii and semi-major axes of {\it Kepler}
  transits by definition for triple-transit systems. Planetary radii show an
  equally good match for double-transit systems, while the single-transit
  systems have significantly larger planets in the {\it Kepler} data.
  Single-transit and double-transit systems have observed planets slightly
  further from the star than the synthetic planets.}
  \label{f:Rp_a_dist}
\end{figure*}

The radius and semi-major axis distribution of the synthetic transits are
compared to the observed transits in \Fig{f:Rp_a_dist}. We focus on a mutual
inclination parameter of $\beta=5^\circ$ which approximately matches the ratio
of double-transit to triple-transit systems in the {\it Kepler} data. The radii
of planets in our observed triple-transit systems in \Fig{f:Rp_a_dist} match by
definition the observed distribution of planet radii in the template
triple-planet systems. A similarly good correspondence is found in the
double-transit systems. This way the triple-planet systems can both produce the
correct relative number of double-to-triple transits and the at the same time
reproduce the size distribution of the planets in double-transit and
triple-transit systems.

However, planets in single-transit systems seen in the {\it Kepler} data
include a much larger fraction of planets having larger radii (between the
radii of Neptune and Jupiter). The synthetic systems almost entirely lack
planets with such large sizes. In addition, the number of single-transit
systems produced from the triple-planet templates is about a factor of three
lower than that observed. A higher number of single-transit systems could in
principle be produced by choosing a higher mutual inclination parameter
$\beta\approx8^\circ$ in \Fig{f:n2n3_n1n3}, however this would not produce a
good match to the observed distribution of planetary radii, as evidenced in
\Fig{f:Rp_a_dist}.

\Fig{f:Rp_a_dist} also shows the distribution of semi-major axis of the
synthetic planets and the {\it Kepler} data. Again there is a perfect match, by
definition, for triple-transit systems, whereas single-transit and
double-transit systems have planets that are on the average a bit further from
the star in the {\it Kepler} data. This discrepancy is nevertheless minor.

\section{Properties of a Second Population}
\label{s:second}

\begin{figure}
  \begin{center}
    \includegraphics[width=0.5\linewidth]{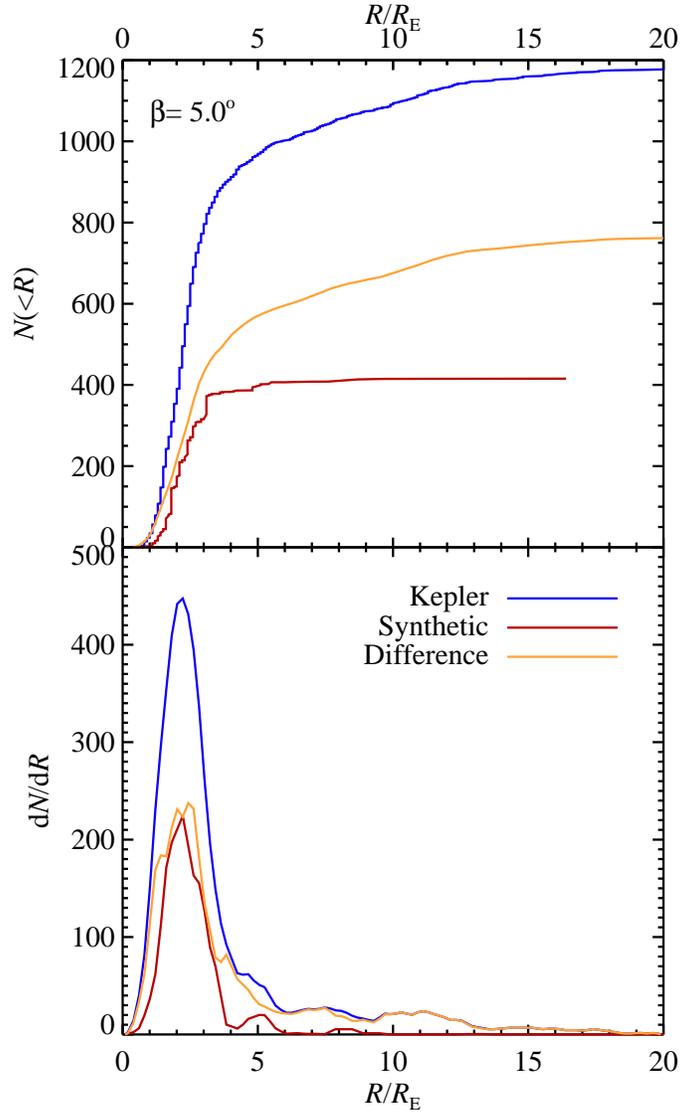}
  \end{center}
  \caption{Cumulative distribution function (top panel) and probability
  distribution function (bottom panel) of the absolute number of observed
  planets (blue) and synthetic planets (red) in single-transit systems versus
  their radius. The difference between observed and synthetic transits is shown
  in yellow. Approximately 750 single-transit planets are missing in the
  synthetic population, which only reproduces 1/3 of the actual number of
  planets in single-transit systems. Approximately 500 of the missing planets
  are small (smaller than four Earth radii) while 250 are large (larger than
  four Earth radii).}
  \label{f:Rp_dist_diff_1t}
\end{figure}

The inability for the template triple-planet systems to explain simultaneously
the double-transit and single-transit systems in the {\it Kepler} data implies
that there is a dichotomy in the underlying planetary systems. On the one hand
there is a population of triple-planet systems with small planets which
faithfully reproduces all triple-transit and double-transit systems, but only
1/3 of the single-transit systems. An additional second population of planetary
systems, distinct from the {\it Kepler} triple-transit systems, produces the
remaining single-transit systems. The cumulative distribution and probability
distribution functions of the planetary radii for observed and synthetic
single-transit systems, in terms of the absolute number of planets, are shown
in \Fig{f:Rp_dist_diff_1t}. We choose a mutual inclination parameter of
$\beta=5^\circ$ which matches the number of double-transit systems in the {\it
Kepler} data. The synthetic data misses around 750 single-transit systems. Of
these approximately 500 are small (smaller than four Earth radii) while 250 are
large (larger than four Earth radii).

More insight into the second population of planets needed to produce the
missing single-transit systems is gained by considering the number of
single-transit, double-transit and triple-transit systems as we increase the
mutual inclination parameter $\beta$. We maintain the triple-planet systems as
templates for illustration, although we show in the next section that unstable
triple-planet systems evolve to stability by having two or more planets collide.

\begin{figure}
  \begin{center}
    \includegraphics[width=\linewidth]{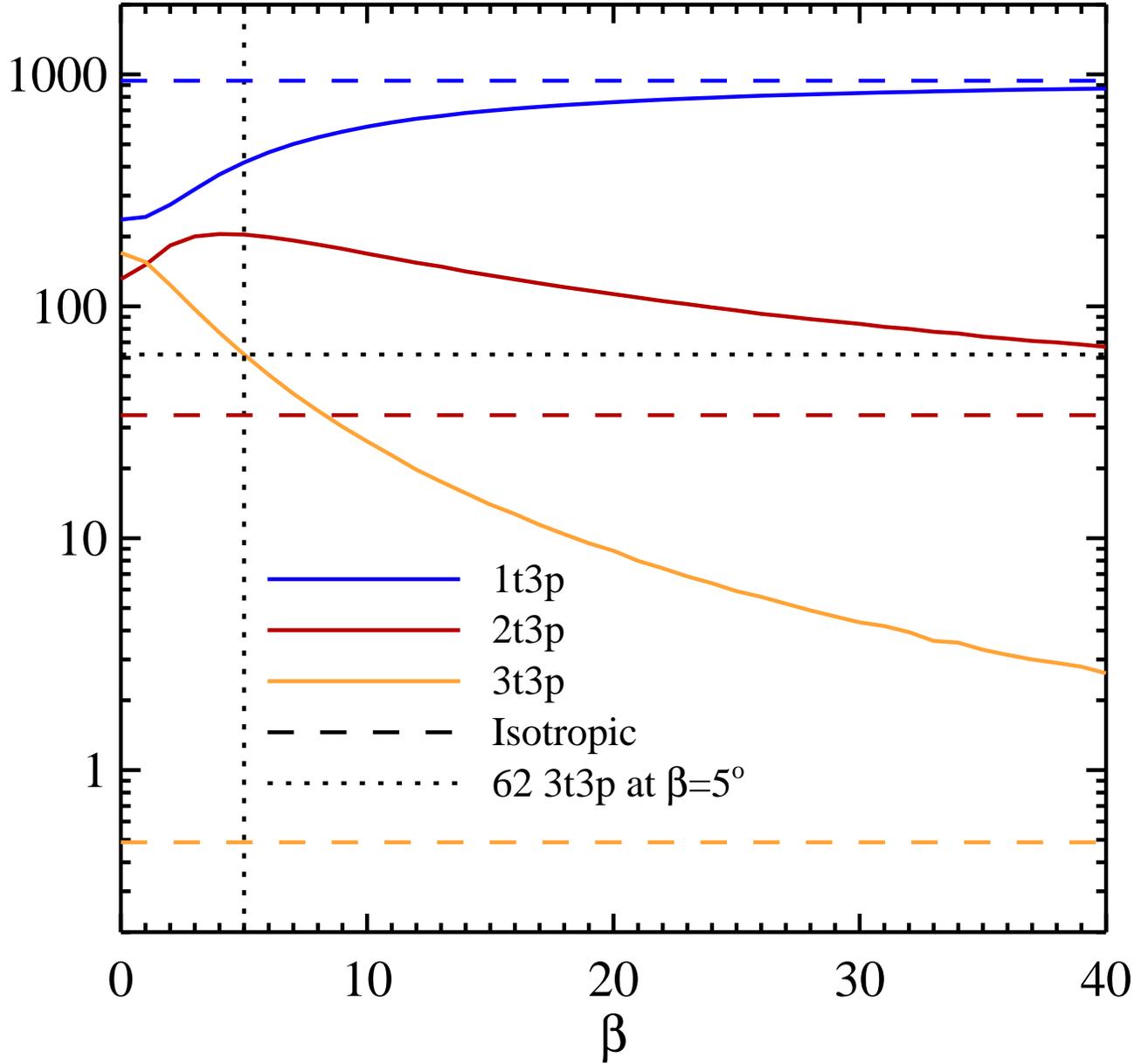}
  \end{center}
  \caption{Number of synthetic single-transit, double-transit and
  triple-transit systems versus mutual inclination parameter $\beta$,
  normalised to match the number of observed triple-transit systems at
  $\beta=5^\circ$. The isotropic limits are indicated with dashed lines.
  Triple-transit systems are quickly suppressed when increasing $\beta$, but
  almost isotropic inclinations are needed to suppress double-transit systems.
  Hence is is difficult for a second population of highly inclined
  triple-planet systems to avoid overproducing (massive) double-transit
  systems.}
  \label{f:n1t_n2t_n3t_beta}
\end{figure}

In \Fig{f:n1t_n2t_n3t_beta} we show the transit number for a mutual inclination
parameter up to 40$^\circ$. We maintain the normalisation that the 62
triple-transit systems found in the {\it Kepler} 16-months data after applying
selection criteria must be produced at the mutual inclination parameter
$\beta'=5^\circ$. The transit numbers $N_i$ are thus defined as
\begin{equation}
  N_i(\beta) = N_3(\beta') \frac{P_i(\beta)}{P_3(\beta')}
\end{equation}
with $N_3(\beta')=62$ to match {\it Kepler} observations. Here the $P_i$ are
the mean $i$-transit probabilities per system, with each system weighted by its
triple-transit probability at $\beta=\beta'$. The normalisation with $N_3$ and
$P_3$ at $\beta=\beta'$ implies that the population of planetary systems with
mutual inclination angle $\beta$ formed with $\beta=\beta'$, just like the flat
population discussed above, and then later heated up to have a higher mutual
inclination parameter of $\beta>\beta'$. The normalisation further implies that
the equation for the transit number can be simplified as
\begin{equation}
  N_i(\beta) = N_3' \frac{\sum_j
  P_{ij} W'_j}{\sum_j W'_j} \frac{\sum_j W'_j}{\sum_j P'_{3j} W_j'} = \sum_j
  P_{ij} W_j' \, ,
\end{equation}
with the short hand notation $N_3' \equiv N_3(\beta')$ and $W_j'=W_j(\beta')$.
\Fig{f:n1t_n2t_n3t_beta} shows that the number of triple-transit systems falls
rapidly with increasing $\beta$ and is unimportant (less than 10\% of the value
at $\beta=0^\circ$) at mutual inclinations above $\beta\approx12^\circ$. The
number of double-transit systems falls much more slowly.

In the isotropic limit where planetary orbits are independent we have
\begin{eqnarray}
  N_1 &=& \sum_j \frac{p_{1j}+p_{2j}+p_{3j}}{P_{3j}(\beta')} - N_2 - N_3 \, , \\
  N_2 &=& \sum_j \frac{p_{1j} p_{2j} (1-p_{3j})}{P_{3j}(\beta')} \nonumber \\
  && + \sum_j \frac{p_{1j}(1-p_{2j}) p_{3j}}{P_{3j}(\beta')} \nonumber \\
  && + \sum_j \frac{(1-p_{1j}) p_{2j} p_{3j}}{P_{3j}(\beta')} \, , \\
  N_3 &=& \sum_j \frac{p_{1j} p_{2j} p_{3j}}{P_{3j}(\beta')}\, . 
\end{eqnarray}
These numbers can be calculated from the output of the synthetic transit
observations. The isotropic limits are shown in dashed lines in
\Fig{f:n1t_n2t_n3t_beta}. In this limit there are 937 single-transit systems,
34 double-transit systems and 0.5 triple-transit systems. At $\beta=40^\circ$
there are nevertheless still 67 double-transit systems, while the number of
single-transit systems has risen to almost 900. In order to produce the 750
missing single-transit systems from a population of triple-planet systems of
high mutual inclination, a mutual inclination parameter of at least 20$^\circ$
would be needed. This high opening angle would as a side effect produce more
than 100 double-transit systems additional to the ones produced from the
low-mass triple systems, enough to push the number of synthetic double-transit
systems significantly beyond the observed number. These extra double-transit
systems would furthermore have the characteristic large planetary radii of the
single-transit systems, in conflict with the excellent radius match that we
find between observed double-transit systems and synthetic double-transits from
the {\it Kepler} triple-systems. Hence it is very difficult for a population of
mutually inclined, massive planets in triple systems to explain the observed
dichotomy between systems showing one transit and systems showing two or three
transits.

In the following we explore the hypothesis that the second population
constituted the upper end of a continuous distribution of planetary birth
masses, and that these systems of massive planets became unstable to
planet-planet interaction and lost one planet by collision or ejection, turning
into double-planet systems with a moderately high inclination between the
planetary orbits. A higher mutual inclination will favour observations of
single transits over observations of double transits, and this way a second
population of double-planet systems can contribute to the missing
single-transit systems without polluting the double-transit systems which are
already produced from low-mass triple-planet systems.

For our hypothesis it is not important whether the massive planets formed in
situ close to their host stars or whether they formed further out and migrated
in due to gravitational torques from the gaseous protoplanetary disk. The
nebula from which the solar system planets formed did not contain enough mass
close to the star to form the solid cores of multiple gas giants there
\citep{Hayashi1981}, although this property is a direct consequence of the
observed lack of massive planets in sub-AU orbits in the solar system.
Extrasolar protoplanetary disks are observed to contain a wide range of masses,
up to 10 times higher than the minimum mass solar nebula, although such massive
disks are relatively rare \citep{AndrewsWilliams2005}. Migration of massive
planets occurs on the viscous time-scale of the protoplanetary disk
\citep{Lin+etal1996}, on the order of a million years, but still much shorter
than the relevant time-scale for planetary instability that we find
\S\ref{s:massboost}. Hence from the view of planetary instability in situ
formation and migration both happen close to instantaneously.

\section{Instability of a High-Mass Second Population}
\label{s:stability}

A possible way to get a moderately or highly inclined population of massive
planets, as discussed in the previous section, is through the intrinsic
instability of planetary systems. If planetary systems are born with a range of
characteristic masses, depending for example on the mass or metallicity of the
protoplanetary disk, then the systems of higher mass can be pushed over the
instability limit for planet-planet interaction. Planet-planet interactions
have been studied in a wide range of contexts, e.g.\ by intrinsic instability
after gas accretion \citep{WeidenschillingMarzari1996,RasioFord1996}, in
connection with scattering of planetesimal belts \citep{Raymond+etal2009} and
after perturbation by a passing star
\citep{ZakamskaTremaine2004,Malmberg+etal2011}.

Long-term integration of triple systems shows that the separation in terms of
mutual Hill radii is an important parameter to determine the stability of
planetary systems \citep{Chambers+etal1996,SmithLissauer2009}. The mutual Hill
radius of two planets of mass $M_1$ and $M_2$ and semi-major axis $a_1$ and
$a_2$ can be defined as
\begin{equation}
  R_{\rm H} = \left( \frac{M_1+M_2}{3 M_\star} \right)^{1/3} \frac{a_1+a_2}{2}
  \, .
\end{equation}
We choose to define the Hill radius at the mean of the planetary semi-major
axes, following convention in the literature, although we note that this
definition does not converge towards the Hill radius of a single massive planet
orbited by a massless test particle in the limit $M_2\rightarrow 0$. The motion
of two planets on circular orbits is stable to their mutual gravity when the
relative separation $\Delta = (a_2-a_1)/R_{\rm H} > 2 \sqrt{3}$
\citep{Gladman1993}. Triple-planet systems do not display a similar abrupt
transition from instability to stability. The logarithm of the time-scale for
instability increases approximately linearly, $\log (t/{\rm Gyr}) \approx b
\Delta + c$, as the mean mutual separation $\Delta$ of the inner and outer
planet pairs is increased. For systems of three low-mass planets,
\cite{Chambers+etal1996} find $b\approx 1.2$ and $c\approx-1.7$. Stability over
at least $10^9$ years then implies $\Delta > 9$.

\begin{figure}
  \begin{center}
    \includegraphics[width=\linewidth]{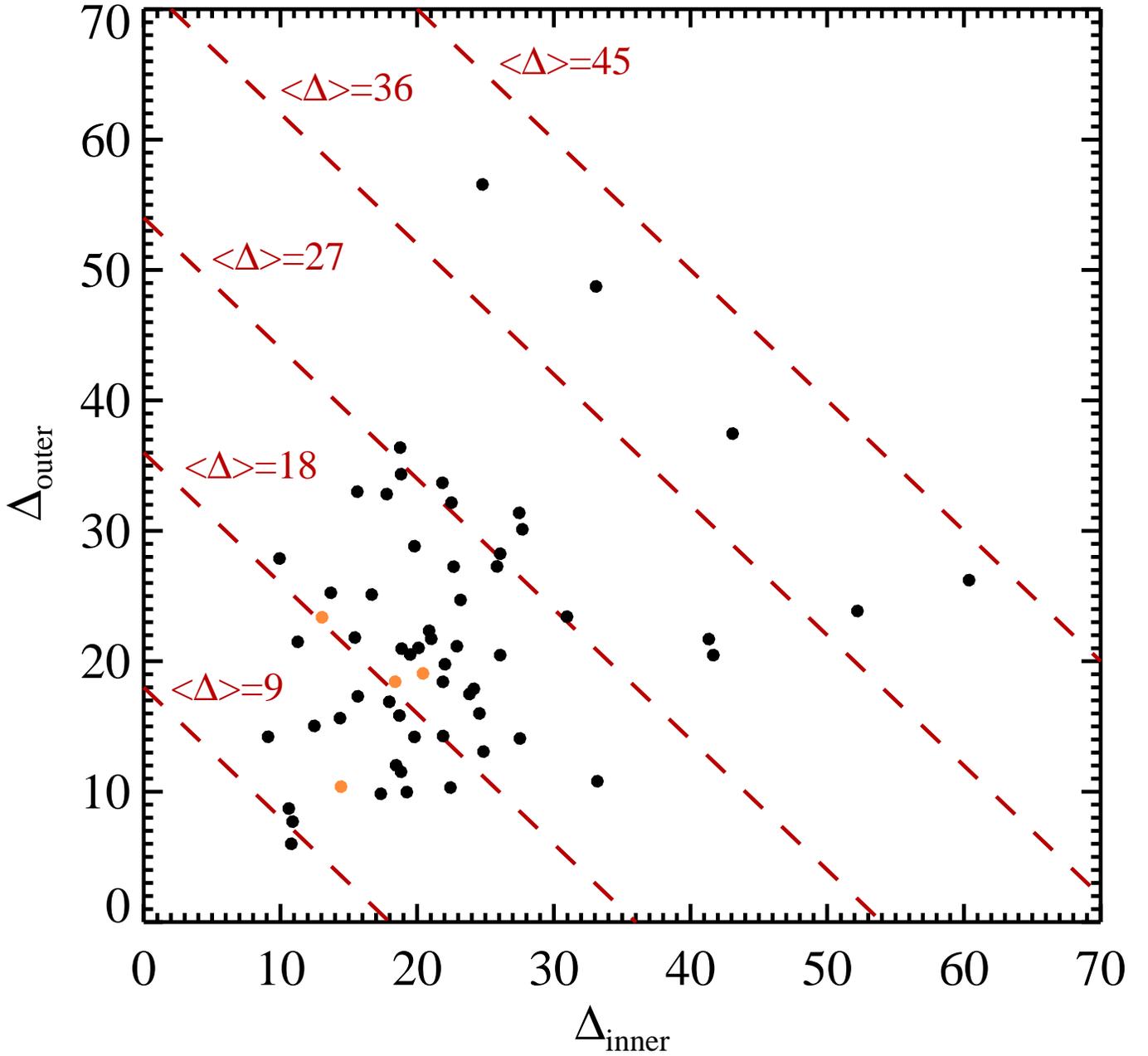}
  \end{center}
  \caption{The separation, measured in mutual Hill radii, of planet neighbours
  in triple-planet systems. The triple-planet stability criterion for
  Earth-mass planets -- a mean Hill separation of 9 -- is given by a dashed red
  line for the generic planet mass based on an approximate mass-radius
  relationship. Mutual Hill radii of [18,27,36,45], corresponding to $\langle
  \Delta \rangle = 9$ for mass boosts of [8,27,64,81], are also indicated. The
  systems that are evolved dynamically in \S\ref{s:stability} are shown in
  orange.}
  \label{f:delta_triple}
\end{figure}

In \Fig{f:delta_triple} we show the mutual separation of planet neighbours, in
units of their mutual Hill radii, for the {\it Kepler} triple-transit systems
(after applying our selection criteria). The masses are based on a simple
mass-radius relationship proposed by \cite{TremaineDong2011} to fit the planets
in the solar system as well as the transiting exoplanets with known masses,
\begin{eqnarray}
  \log (R/R_{\rm J}) &=& 0.087 + 0.141 \log (M/M_{\rm J}) \nonumber \\
  & & \quad - 0.171 (\log M/M_{\rm J})^2 \, . \label{eq:massrad}
\end{eqnarray}
The radius-mass relationship increases monotonically up to planets of around
the mass of Jupiter, and then turns over with a relatively constant radius for
higher-mass planets. The high-radius branch can not be easily inverted to
obtain the mass, but the triple-planet systems in the {\it Kepler} data contain
very few planets above 10 Earth radii, so the degenerate mass-radius
relationship is not an issue for our calculations. All systems in
\Fig{f:delta_triple} are stable or at the edge of stability over $10^9$ years
when using their nominal masses. The maximum mean Hill separation is
approximately 45, but 89\% of the systems have $\langle \Delta \rangle<30$. The
lack of systems of high Hill separation is likely partially an observational
bias, since widely spaced systems are less likely to be observed in triple
transit.

Higher-mass versions of the {\it Kepler} triple-planet templates would, with a
sufficient mass boost, be unstable to planet-planet scattering. These systems
would not show up in the triple-transit systems observed by the {\it Kepler}
mission because they have lost one or more planets by ejection or collisions
(between planets or with the host star), leaving behind excited remnants of the
original systems. We investigate the stability of mass-boosted triple-transit
{\it Kepler} systems by performing $N$-body simulations using the orbital
dynamics code MERCURY \citep{Chambers1999}. We define the mass boost as $MB =
M_i/M_{i0}$ where $M_{i0}$ is the mass of planet $i$ from an approximate
mass-radius relationship and $M_i$ is the boosted mass. The MERCURY code uses a
symplectic integrator to achieve conservation of linear and angular momentum
and is thus an optimal tool for evolving planetary systems for a high number of
orbital periods.

\subsection{Mass-Boosted Systems}
\label{s:massboost}

\begin{figure}
  \begin{center}
    \includegraphics[width=\linewidth]{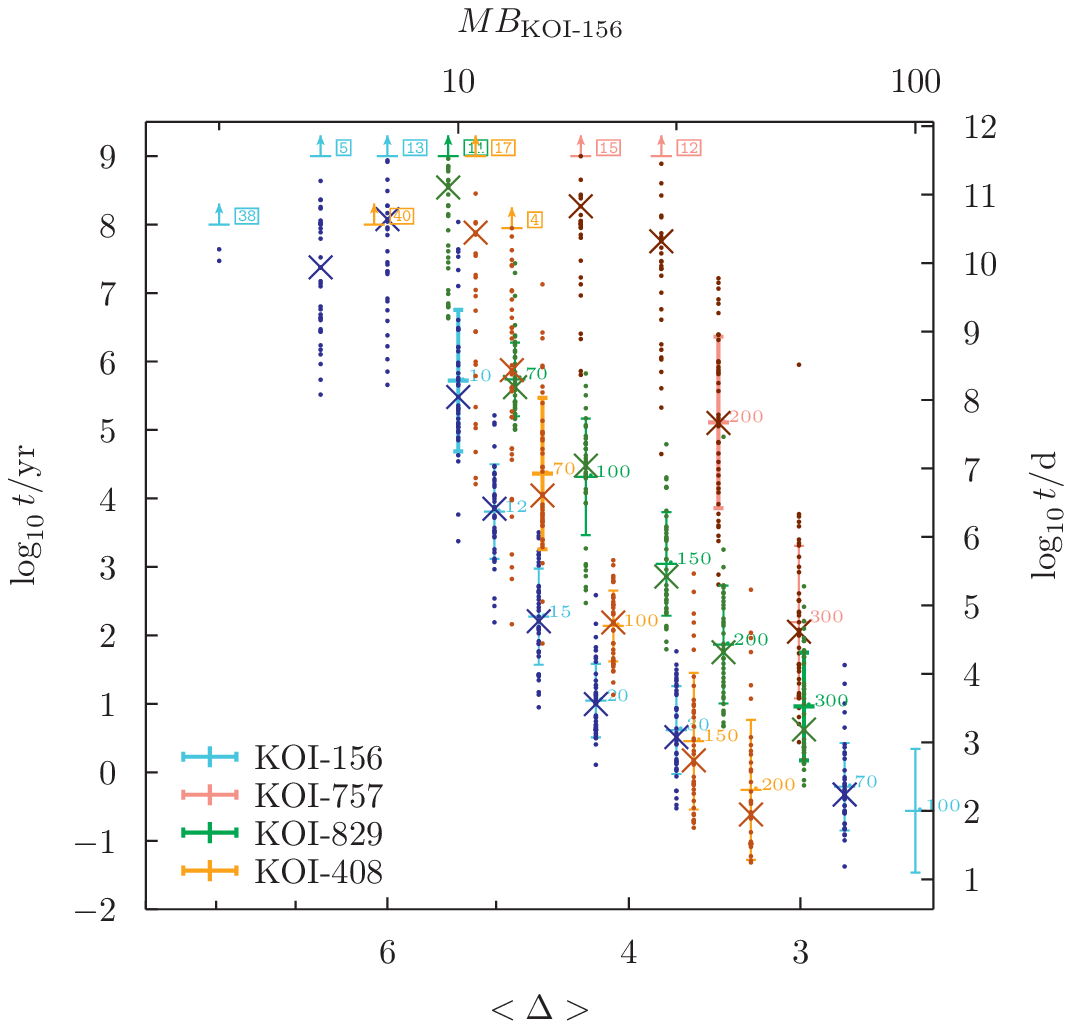}
  \end{center}
  \caption{Time-scale for a close encounter between two planets to occur in
  mass-boosted triple-planet systems, as a function of the mean initial
  separation in units of mutual Hill radii, of four representative
  triple-planet systems (KOI-156, KOI-757, KOI-829 and KOI-408). Dots show the
  results of different initial realisations of the planet orbits for a given
  $\langle \Delta \rangle$, with big crosses indicating the median instability
  time-scale and pluses the mean (the bar shows the standard deviation).
  Individual mass boosts are printed next to the mean, while the top axis shows
  the mass boost for KOI-156. The number of implementations of each system that
  are stable for more than either $10^8$ or $10^9$ years are shown at the top
  of the plot next to an upwards-pointing arrow. Most systems need a mean Hill
  radius separation $\langle \Delta \rangle < 6$ to become unstable within
  $10^{10}$ years.}
  \label{f:timescale}
\end{figure}

The triple-planet systems observed by the {\it Kepler} mission have been
carefully checked for dynamical instability using the nominal planetary masses
\citep{Lissauer+etal2011b,Fabrycky+etal2012}. Here we report results of orbital
integration of mass-boosted counterparts to the observed triple-planet systems.
In \Fig{f:timescale} we plot the results of $N$-body simulations of four
representative systems from the template triple-planet systems -- KOI-156,
KOI-757, KOI-829 and KOI-408. These systems are indicated with orange dots in
\Fig{f:delta_triple}. We run each system for a number of mass boosts and
monitor the time for the first close encounter between two planets. We consider
this the relevant time-scale for instability of the planetary system. Each
system is run for 40 random initialisations of the planetary orbits. The
different implementations show instability at different times, but the median
time-scale increases monotonously with the mean Hill separation between planet
neighbours, $\langle \Delta \rangle$. We stop the simulations either after
$10^8$ years or after $10^9$ years. Results can be extrapolated to $10^{10}$
years, which we take as an upper limit for the age of the {\it Kepler} target
stars. Although the extrapolation has a large scatter between the individual
systems, as noted by \cite{DuncanLissauer1997} who performed similar
mass-boosted simulations of the Uranian moon system, the overall statistical
trend for the instability time-scale to increase rapidly with increased Hill
separation is robust.

\begin{figure}
  \begin{center}
    \includegraphics[width=\linewidth]{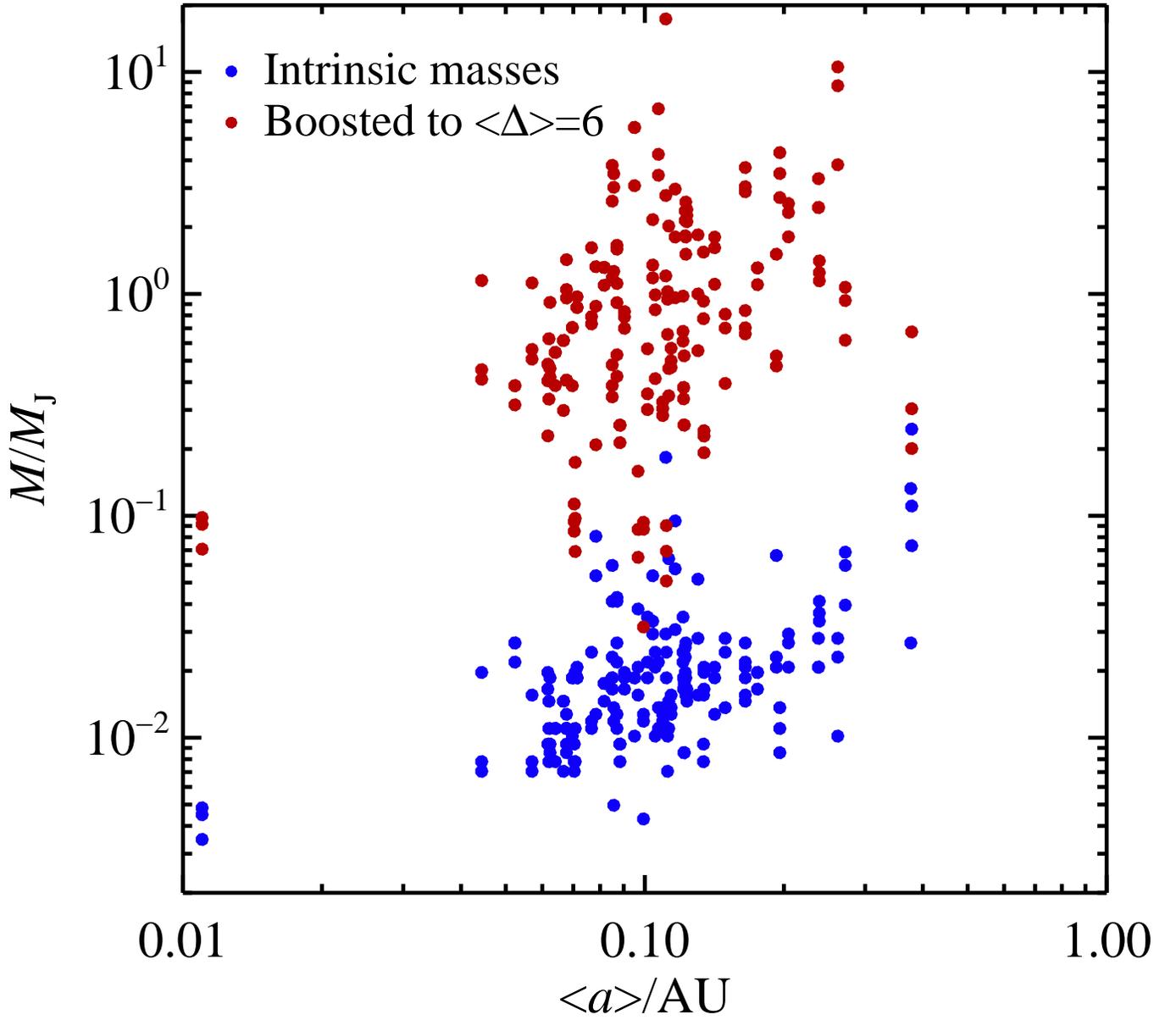}
  \end{center}
  \caption{Masses of planets in the triple-planet template systems as functions
  of the mean semi-major axis of planets in the system. The blue points show
  the intrinsic masses based on an approximate mass-radius relationship, while
  the red points show the masses after boosting each system to expected
  instability at a mean Hill separation of $\langle \Delta \rangle = 6$. The
  boosted planetary masses are between 0.1 and 10 Jupiter masses, with the
  majority of the planets between 0.3 and 3 Jupiter masses.}
  \label{f:mass_massboost}
\end{figure}

\Fig{f:timescale} shows that the systems need $\langle\Delta\rangle\lesssim6$
in order to become unstable within $10^{10}$ years. This limit is much lower
than found by \cite{Chambers+etal1996} from integrations of triple systems of
objects with masses comparable to planetary embryos ($M \approx 0.03
M_\oplus$). As noted in \cite{Chambers+etal1996}, increasing the characteristic
planetary mass, while maintaining the mean Hill separation, leads to
systematically longer instability time-scales. In \Fig{f:mass_massboost} we
show the intrinsic masses of the planets in the triple-planet template systems,
based on the approximate mass-radius relationship in \Eq{eq:massrad}, as a
function of the mean of the semi-major axis. We also indicate the masses of the
systems when boosted to expected instability at a mean Hill separation of
$\langle \Delta \rangle \approx 6$. The mass-boosted systems have
characteristic planetary masses between 0.1 and 10 Jupiter masses, with the
majority of the planets between 0.3 and 3 Jupiter masses.

There are three possible (non-exclusive) outcomes when a planetary system
becomes unstable: 1) the system may eject one or more planets; 2) two or more
planets may collide; or 3) one or more planets may be scattered into the host
star. The relative frequency of these three outcomes is a function of the ratio
of the planetary orbital speeds to their surface escape speeds, the square of
which is often referred to as the Safronov number
\citep[e.g.][]{SafronovZvjagina1969,BinneyTremaine2008}
\begin{equation}
  \theta_{\rm S} = {\frac{1}{2}}{\frac{v_{\rm esc}^2}{v_{\rm orb}^2}} \, .
\end{equation}
Note that this definition is the inverse of the one used in
\cite{FordRasio2008}. In our definition, for planetary systems having $\theta
\gg 1$, planet-planet scatterings can result in the ejection of planets.
However for systems where $\theta < 1$, collisions will be common as a
significant deflection would only occur for close encounters (between two point
masses) with minimum separations smaller than the planetary radii. The surface
escape speed for a Jupiter-like planet is about 60 km/s, whereas the orbital
speed at 1 AU around a solar-mass star is 32 km/s; and at 0.1 AU about 100
km/s. Hence unstable planetary systems containing planets of Jovian mass will
eject planets if their orbits have semi-major axes around or above 1 AU, whilst
systems containing planets or tighter orbits around 0.1 AU will undergo
planetary collisions (leading to mergers) or have planets scattered into their
host stars. Collisions are thus expected to be the most common event for the
mass-boosted planets in the {\it Kepler} data.

\begin{figure}
  \begin{center}
    \includegraphics[width=0.5\linewidth]{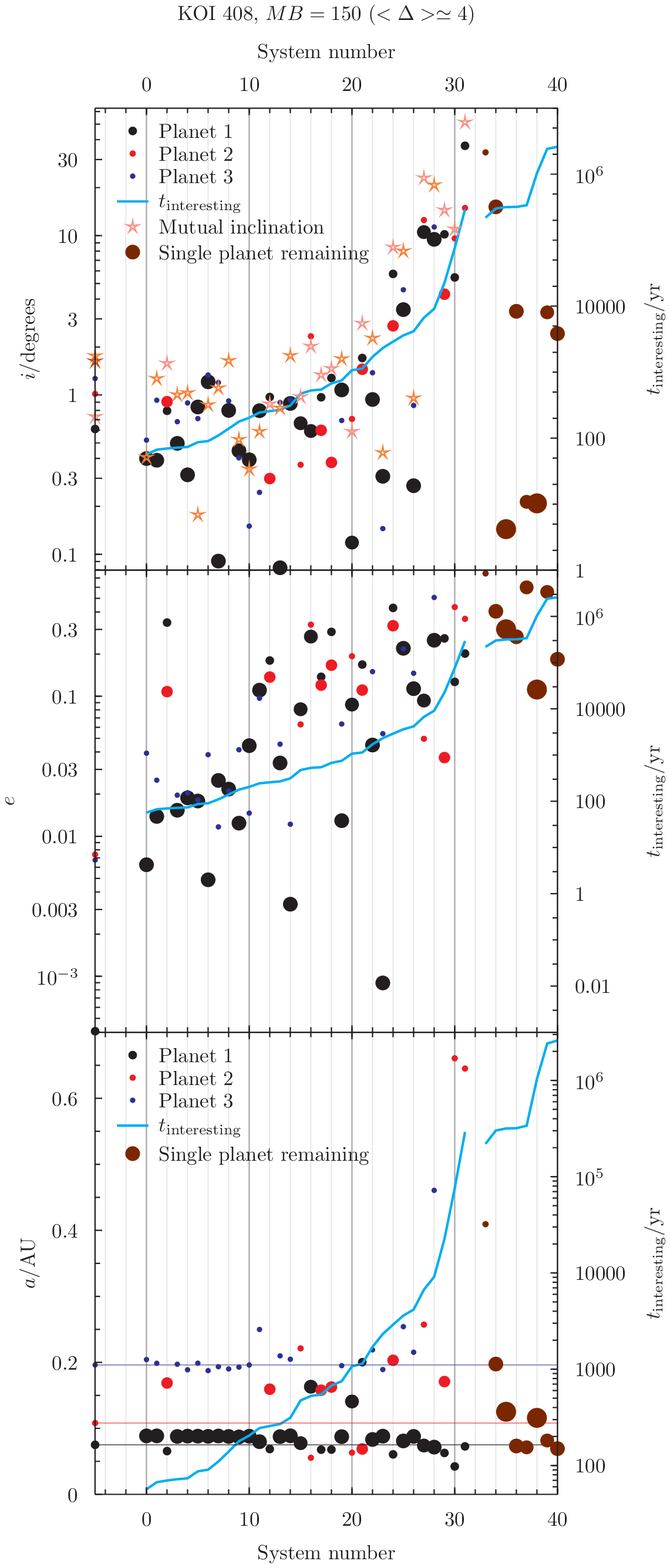}
  \end{center}
  \caption{Result of orbital integration of KOI-408. The bottom panel shows the
  semi-major axis after planetary instability (left axis, planets
  indicated with circle radius proportional to planetary radius) and the
  time-scale for the first ejection or collision (right axis, turquoise
  curve). The $x$-axis shows the results of forty initial
  representations of the mass-boosted system, sorted by increasing time for
  instability. An empty column at position 31 is used to separate systems with
  two remaining planets from systems with one remaining planet. Typically the
  middle planet collides with the inner planet, or in cases where the system
  falls apart slowly, all three planets merge. The middle and top panels show
  the eccentricities and mutual inclinations of the post-encounter systems. The
  resulting eccentricity is typically $\sim$$0.1$, while the mutual inclination
  ranges between 1 and 30 degrees.}
  \label{f:final}
\end{figure}

In \Fig{f:final} we show the result of integrating KOI-408 with a mass boost of
150 (corresponding to a mean Hill separation $\langle\Delta\rangle=4$). The
bottom panel shows the semi-major axes of the planets after planetary
instability has relaxed the system to a stable configuration. The $x$-axis
shows the results of forty initial representations of the mass-boosted system,
sorted by increasing time for instability. The most common result is that the
inner and middle planets collide. Only in systems having a  long instability
time-scale do we find collisions involving all three planets. The middle and
top plots show the associated eccentricities and mutual inclinations.
Eccentricities up to 0.3 are common, while the mutual inclination varies from a
few degrees to up to 30 degrees. This way the typical inclination is
increased significantly by planet-planet scattering, so that the resulting
double-planet systems produce many observed single transits. High
eccentricities are also common among giant exoplanets observed in radial
velocity surveys \citep{UdrySantos2007} and our results show that planet-planet
scattering {\it in situ} in the inner AU can cause both high eccentricities and
inclinations.

\begin{figure}
  \begin{center}
    \includegraphics[width=\linewidth]{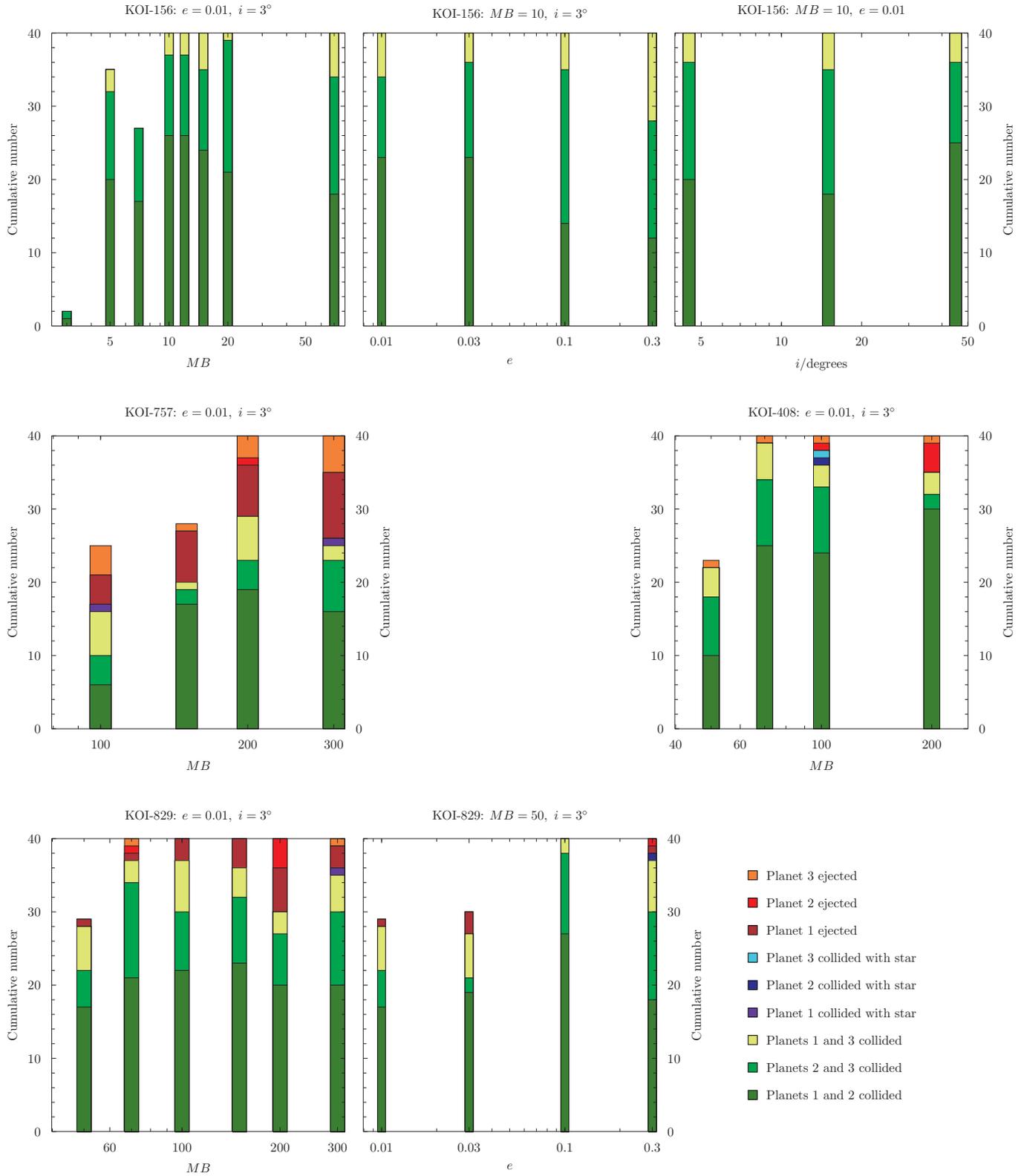}
  \end{center}
  \caption{Outcome of planetary instability of KOI-156, KOI-757, KOI-408 and
  KOI-829 for different mass boosts. KOI-156 was furthermore run for four
  different values of the eccentricity and three values for the inclination and
  KOI-829 for two values of the eccentricity. Each parameter choice was run for
  40 different random initialisations. Green denotes planet-planet collisions,
  blue denotes planet-star collisions and red/brown ejections. Planet-planet
  collisions are by far the most common event due to the proximity to the
  star. Ejections are less common, but their frequency increase with
  increasing mass boost. Collisions with the star are very rare.}
  \label{f:colltype}
\end{figure}

\Fig{f:colltype} shows the fate of planets in all four dynamically evolved
systems as a function of the mass boost (and the eccentricity and inclination
for selected systems). Green denotes planet-planet collisions, blue denotes
planet-star collisions and red ejections. The most common events are
planet-planet collisions, as expected for these close-in systems. Ejections are
far less common, but their frequency increases with increasing mass boost.
Collisions with the star are very rare.

\begin{figure}
  \begin{center}
    \includegraphics[width=\linewidth]{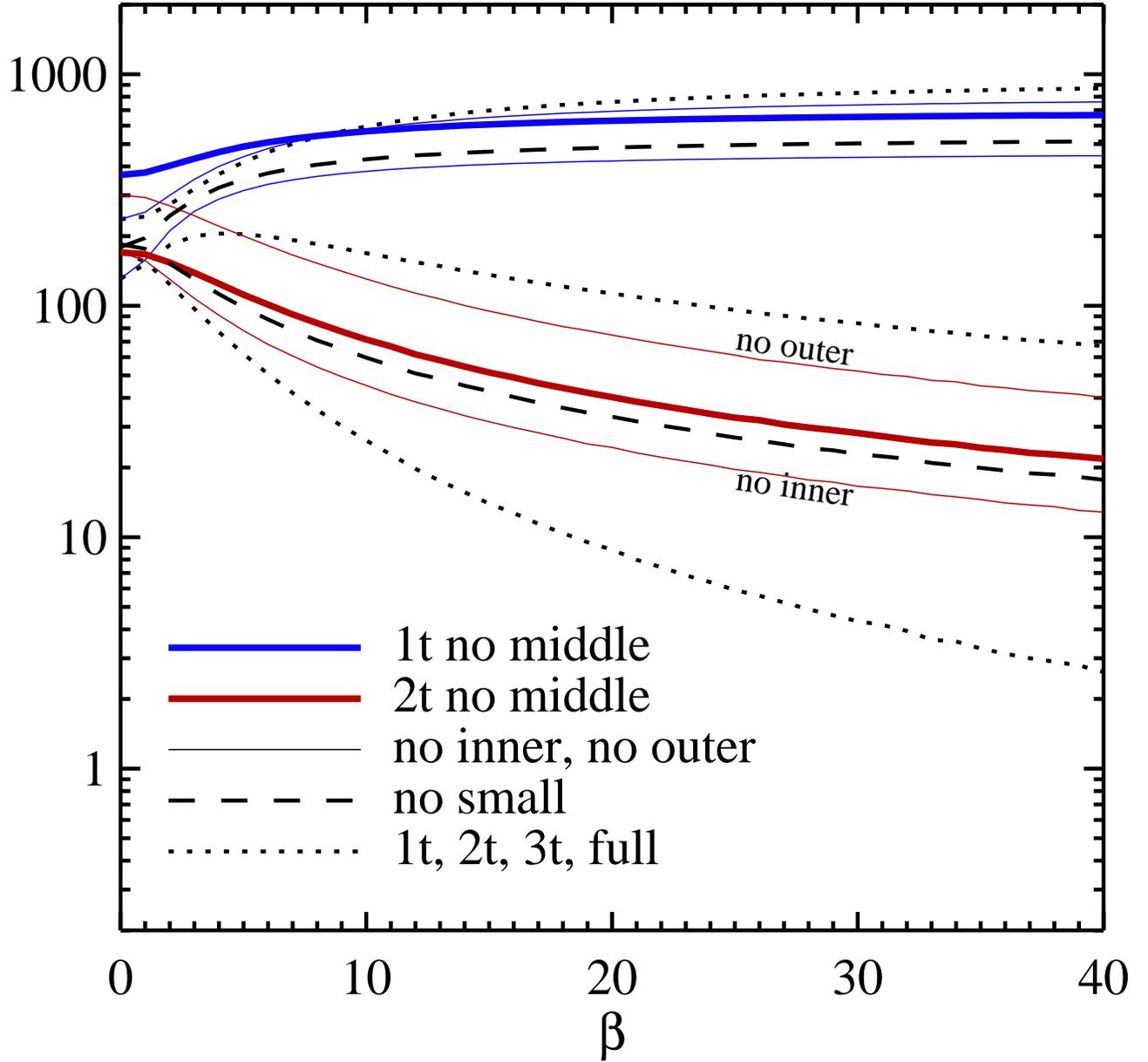}
  \end{center}
  \caption{The transit numbers after removing a planet by planet-planet
  collision, collision with the star or ejection. The blue lines show the
  number of single-transit systems, when the middle (full line) or inner/outer
  planet (thin line) is removed. The red lines show the number of
  double-transit systems. The dashed line shows the result of removing the
  smallest planet in the system. Triple-transit systems vanish. The number of
  double-transit systems is reduced, compared to using the original
  triple-planet systems, relatively more than the number of single-transit
  systems.}
  \label{f:n1t_n2t_reduced_beta}
\end{figure}

The result of planet-planet scattering is most often a collision between two
planets, reducing the number of planets from three to two. Taking this result
as a recipe, we can create the double-planet systems arising from dynamical
instability of triple systems by removing a planet from the template
triple-planet systems. The probabilities for individual planets to transit
remain unchanged since the (small) effect of the planetary radius is ignored
when calculating the transit probability, although the number of
single-transit, double-transit and triple-transit systems change when a planet
is removed (the latter number obviously falls to zero). The resulting transit
numbers are shown in \Fig{f:n1t_n2t_reduced_beta}. We experiment by removing
either the middle planet or the outer or inner planet. For simplicity we
maintain the original planetary orbits, although a more detailed analysis
should put a planet resulting from a collision on an intermediate orbit between
the two colliding planets. The dashed lines in \Fig{f:n1t_n2t_reduced_beta}
furthermore show the result of removing the smallest planet in the systems.
Triple transits vanish in all these two-planet systems. As the inclination
parameter $\beta$ is increased the number of single transit-transit systems
increases at the cost of double-transit systems.

The removal of a {\it random} planet in each system would reduce the number of
single-transit systems to 2/3 and the number of double-transit systems to 1/3
compared to the template triple-planet systems. Thus we expect that the ratio
of single-transit to double-transit systems will approximately double compared
to the full systems, independently of $\beta$. \Fig{f:n1t_n2t_reduced_beta}
shows that it is possible to observe 600 single-transit systems, with only 10
double-transit systems accompanying, if $\beta$ is moderately high at around 40
degrees. This way the massive double-planet systems resulting from planetary
instability of triple-planet systems can produce a large number of
single-transit systems with massive planets and at the same time avoid
polluting the double-transit systems with planets from the massive population.

\section{Systems with a Distribution of Multiplicities}
\label{s:multiple}

The considerations in the previous section have an inherent problem in that the
majority of the missing planets in single-transit systems are small (smaller
than four Earth radii) and thus would not contribute to making a system
unstable. \Fig{f:mass_massboost} shows that only systems with characteristic
planetary masses between 0.1 and 10 Jupiter masses can undergo dynamical
instability. Planetary instability of triple-planet systems of high birth
masses can in principle result in a population of double-planet systems that
produce the 250 observed single-transit systems containing large planets (more
than four times the radius of the Earth), if the high-mass population
represented approximately 50\% of the number of low-mass triple-planet systems
that have survived intact until today. \Fig{f:delta_triple} shows that there is
a void of planetary systems with mean mutual Hill separation $\langle \Delta
\rangle \lesssim 9$. This corner may have been occupied after planet formation
and become unstable to planet-planet scattering as the gaseous protoplanetary
disk dissipated. Extrapolating the number of systems between $\langle \Delta
\rangle = 9$ and $\langle \Delta \rangle = 18$ and between $\langle \Delta
\rangle = 27$ and $\langle \Delta \rangle = 36$ to $\langle \Delta \rangle < 9$
shows that a significant number of systems could have formed there.

However, the high-mass systems can not explain the surplus of single-transit
systems with smaller planets. Double-transit systems and triple-transit systems
also display a general lack of high-mass planets (see \Fig{f:Rp_P}), a property
that is difficult to reconcile with the hypothesis that some systems are simply
born with high-mass planets, since triple-planet systems could have one or two
massive planets and still be stable over long time-scales.

\begin{figure}
  \begin{center}
    \includegraphics[width=\linewidth]{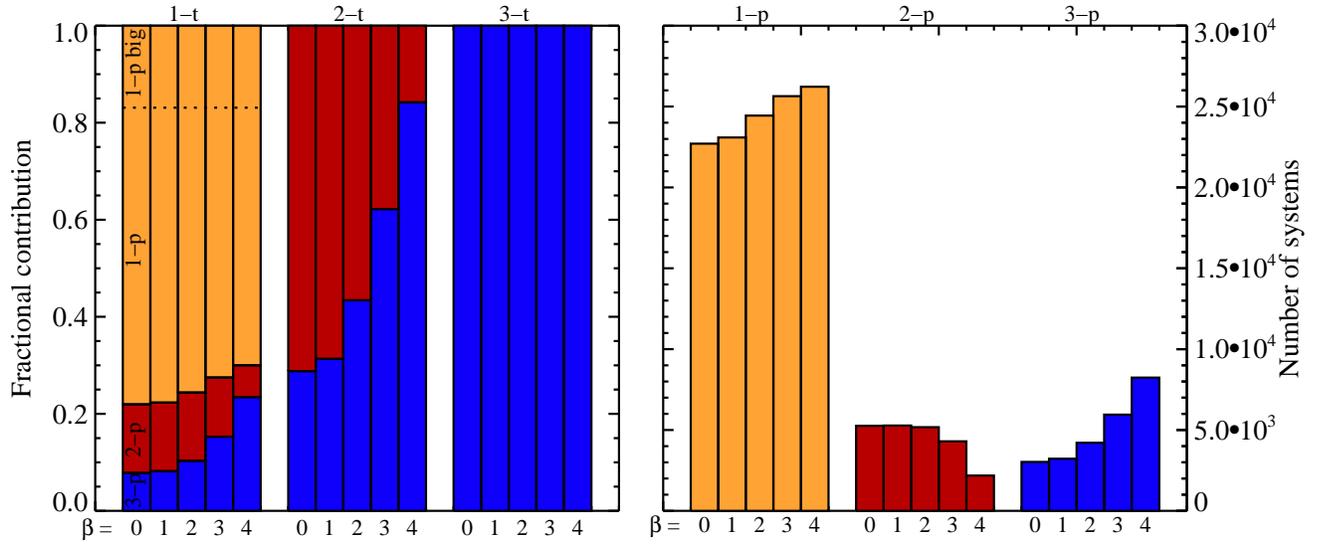}
  \end{center}
  \caption{Fractional contribution to transits (left panel) and number of
  planetary systems (right panel), as functions of mutual inclination parameter
  $\beta$ (horizontal axis) and system multiplicity (yellow bars: single-planet
  systems, red bars: double-planet systems, blue bars: triple-planet systems).
  Here we have relaxed the assumption that all planetary systems are triple and
  allowed for both single-planet, double-planet, and triple-planet systems.
  Increasing $\beta$, the number of triple-planet systems increases, as does
  their contribution to single and double transits. The number of double-planet
  systems decreases slowly as $\beta$ increases, eventually dropping below zero
  for $\beta=5^\circ$ (not shown). The necessary population of large
    single-planet systems is marked with a dotted line.}
  \label{f:multiple_population}
\end{figure}

A route to producing the single-transit systems with small planets without
making too many double-transit and triple-transit systems (which are already
produced from the triple-planet system templates) is to consider additional
components of double-planet and single-planet systems, together with the
triple-planet templates. Choosing a random inclination parameter even smaller
than $\beta=5^\circ$ for the triple-planet systems, the number of
double-transit systems arising from the triple-planet templates can be reduced.
Additional populations of double-planet and single-planet systems must then be
present to produce the remaining single-transit and double-transit systems. In
\Fig{f:multiple_population} we show the resulting system architectures. We make
the assumption that all systems have the same mutual inclination parameter, a
reasonable assumption since we have already shown that the double-transit and
triple-transit systems in the {\it Kepler} data are statistically similar with
regards to planetary sizes and orbits. We can then reconstruct the intrinsic
number of single, double and triple systems by requiring first that the
triple-planet systems must produce all the triple-transit systems, then that
the double-planet systems must produce the double-transit systems that are not
already produced by the triple-planet systems, and finally that the
single-planet systems must produce the single-transit systems not produced by
double-planet and triple-planet systems. We create sparser systems from the
triple-planet templates by removing one or two random planets from each system.

\Fig{f:multiple_population} shows that as the mutual inclination parameter is
varied from $0^\circ$ to $4^\circ$, the number of triple-planet systems
increases from 3,000 to over 8,000. At the same time their contribution to
double-transit and single-transit systems increases. This happens at the
expense of double-planet systems whose number falls from approximately 5,000 to
less than zero at $\beta=5^\circ$. The number of single-planet systems
increases from slightly above 20,000 to almost 27,000 as $\beta$ increases from
$0^\circ$ to $4^\circ$. This implies that a significant fraction, more than
20\%, of the 156,000 {\it Kepler} target stars have detectable planets within
0.5 AU, and that the majority of planet-hosting stars have only a single planet
in this region. We caution, however, that this number relies strongly on the
assumption that rich systems of moderate or high mutual inclination do not
contribute significantly to the observed transits.

Systems of higher multiplicity than three are ignored in the above analysis.
Without performing full synthetic transit observations on such systems, which
is undesirable since their number and statistical significance is low, we can
derive useful limits to their contribution to triple, double and single
transits by considering the case $\beta=0^\circ$. Completely flat systems have
the property that the probability that system $j$ (of multiplicity $I$)
displays exactly $i$ transits is
\begin{eqnarray}
  P_{ij} &=& \frac{R_{\star j}}{a_{ij}} \quad\quad\quad\quad\quad\,\, {\rm
  for} \quad i=I \, , \label{eq:pij_flat_1} \\
  P_{ij} &=& \frac{R_{\star j}}{a_{ij}} - \frac{R_{\star j}}{a_{(i+1)j}} \quad
  {\rm for} \quad i<I \label{eq:pij_flat_2} \, .
\end{eqnarray}
The transit of the outermost planet always implies that all $I$ planets transit
(equation \ref{eq:pij_flat_1}), while the transit of planet $i$ (with $i<I$)
corresponds to exactly $i$ planets transiting only if planet $i+1$ does not
transit (equation \ref{eq:pij_flat_2}).

Analysing this way the 14 quadruple-transit systems which are present in the
{\it Kepler} data after our selection criteria, we find that the intrinsic
population of quadruple-planet systems is around 400, far less than the 3,000
triple-planet systems present at $\beta=0^\circ$ (see
\Fig{f:multiple_population}). These 400 quadruple-planet systems produce 14
quadruple transits, 8 triple-transits, 16 double transits and 30
single-transits, using the two transit-rate equations above. A similar analysis
on the 4 quintuple-transit systems yields an underlying population of 130
quintuple-planet systems, which give rise to 2.3 quadruple-transit systems, 3.4
triple-transit systems, 4.9 double-transit systems and 9.3 single transit
systems. While systems of higher multiplicity than three must be present in the
{\it Kepler} target stars, we can conclude that if the planetary systems are
intrinsically very flat, then their contribution to the total number of
planetary systems as well as to the lower-order transit systems is low.
However, if the intrinsic mutual inclinations are non-zero, then a large
fraction of the observed triple-transit systems may arise from more crowded
planetary systems \citep{RagozzineHolman2010}.

Planetary systems can nevertheless not be made arbitrarily flat, unless the
eccentricities in the system are equally low. Secular oscillations cause
exchange between eccentricity and inclination and hence do not allow very flat
systems. Still, the need for very flat triple-planet systems (as well as some
systems of higher multiplicity), with additional populations of low-mass
single-planet and double-planet systems is appealing in that the abundance of
planetary systems is continuous in the multiplicity.

\section{Summary and Conclusions}
\label{s:conclusions}

This paper sets out to construct underlying planetary system architectures that
explain the systems of single and multiple transiting planets in the {\it
Kepler} data. One of the most robust features of the {\it Kepler} data is that
there is a dichotomy between the planets in systems displaying two or three
transits and the planets in systems displaying only one transit. Double-transit
and triple-transit systems have mostly small (less than four Earth radii)
planets, while the single-transit systems host both small planets and an
additional population of large planets.

The dichotomy between large planets in single systems and small planets in
systems with a wide range of multiplicities was also detected in the transit
synthesis models of \cite{Lissauer+etal2011b} and of \cite{TremaineDong2011},
but we confirm it here using an alternative approach to constructing synthetic
planetary systems. In this approach we assume that the systems of three
transiting planets observed by the {\it Kepler} mission can serve as templates
for physical planetary systems. We thus work close to the original data and do
not have to go through steps of interpretation and analysis of the statistical
properties of planet sizes and orbits to fit these with distribution functions.

We furthermore assume that systems with three transits inherently arise from
physical triple-planet systems. This assumption may in reality not hold, since
quadruple-planet systems and higher contribute to triple transits. However, we
argue that it is reasonable to assume that denser systems are rare and thus
only make minor contributions to observed triple-transit systems. The same
assumption also excludes the isotropic solutions of \cite{TremaineDong2011}
where planets are either in single systems or in extremely packed systems with
almost isotropic orbits.

We explore the underlying hypothesis that triple-planet systems form with a
range of characteristic masses and that the systems of high mass are inherently
unstable. Using $N$-body simulations we find that the most likely outcome of
planetary instability is collision between two of the planets. Collisions are
more likely than ejections because the planetary orbits are so close to the
host stars. The resulting double-planet systems have moderate mutual
inclinations and would produce a number of single transits from large planets.
We find that the instability time-scale is very long unless we use high mass
boosts. Triple-planet systems would need to have characteristic planetary
masses between 0.1 and 10 Jupiter masses in order to produce dynamical
instability within 10 billion years. 

Another generic problem with invoking a second population of massive planets to
explain the observed large planets in single-transit systems is that the
observed double-transit and triple-transit systems generally lack planets above
four Earth radii, while these systems could remain stable with one or even two
gas-giant planets. We consider this a major weakness for the hypothesis that
the large planets in single-transit systems are a result of planetary
instability of the high-end tail of planetary birth sizes.

\begin{figure}
  \begin{center}
    \includegraphics[width=0.8\linewidth]{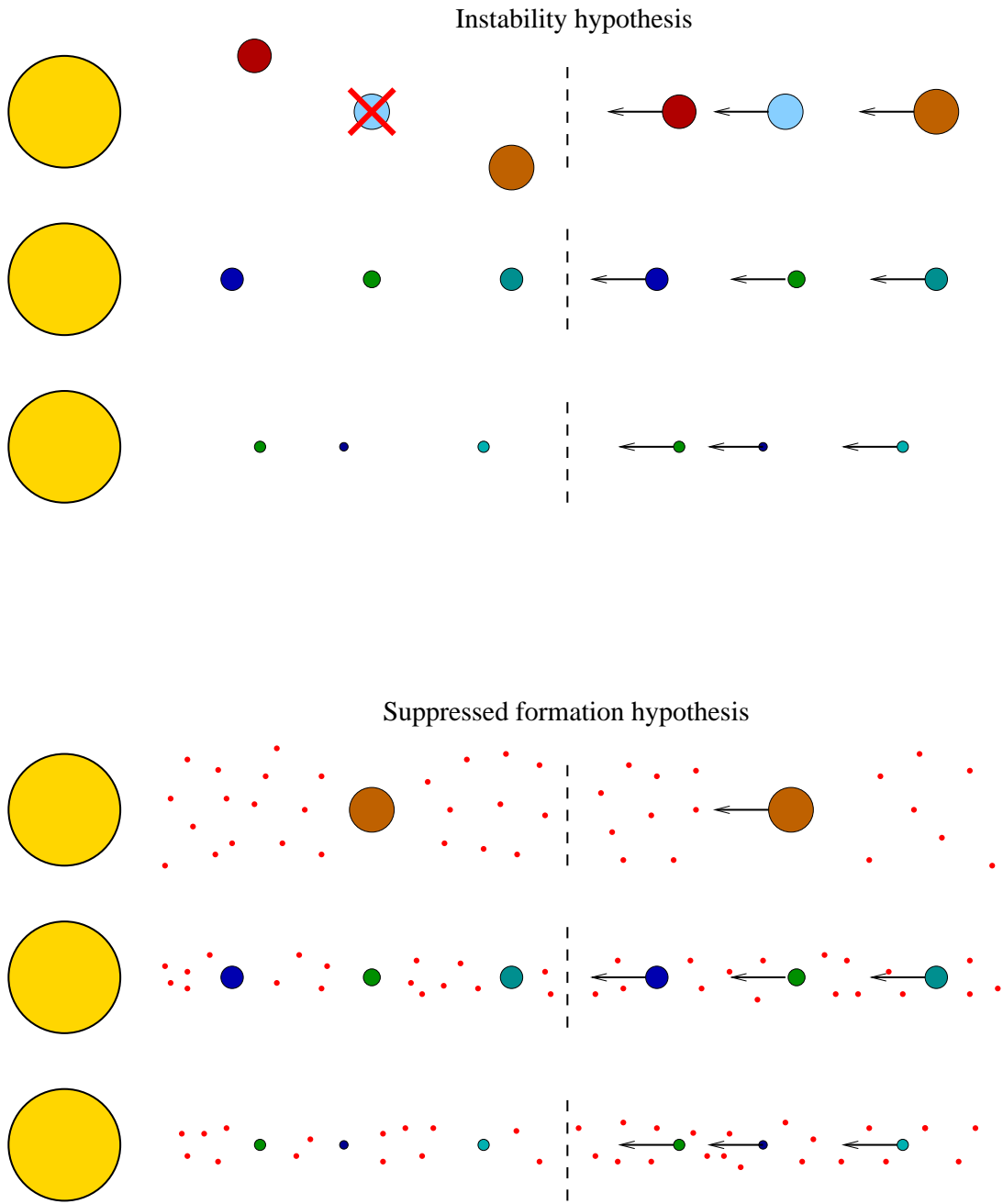}
  \end{center}
  \caption{Sketch of the various situations that can lead to a separate
  population of massive single-planet systems. In the top panel triple-planet
  systems form with various masses, either in situ (left) or further out
  followed by migration (right). The more massive systems are unstable and
  reduce the planet number by planet-planet collisions, leaving two planets of
  moderately high mutual inclination. However, the masses needed to make the
  systems unstable are very high. Instead the formation or migration of a
  gas-giant planet may suppress formation of other (small and large) planets in
  the sub-AU regions of the system, since the gas giant excites high
  eccentricities and inclinations of the planetesimal and embryo population
  (lower panel).}
  \label{f:sketch}
\end{figure}

We find instead that the dichotomy can be seen as the contrast between a
continuous population of almost flat systems with one or more small planets
within 0.5 AU versus a population of inherently single systems with large
planets. In this picture the excess of large planets in single-transit systems
may arise already at the planet formation stage, when the formation or
migration of a massive gas giant in a system suppresses the formation of
additional (small and large) planets \citep[as proposed
by][]{Latham+etal2011}. This scenario is sketched in \Fig{f:sketch}. Such a
mechanism has been put forward to have frustrated the growth of Mars in the
solar system during Jupiter's period of inwards migration
\citep{Walsh+etal2011}, but may have been even more effective in extrasolar
systems where gas giants have migrated to sub-AU orbits.

The {\it Kepler} dichotomy appears to be ultimately controlled by the
metallicity of the host star, as gas-giant planets in sub-AU orbits are mainly
present above a threshold metallicity around the solar value
\citep{Santos+etal2004,FischerValenti2005,Sousa+etal2011}, while Neptune-sized
planets, super-Earths and terrestrial planets are found in the {\it Kepler}
data around stars with a wide range of metallicities \citep{Buchhave+etal2012}.

\acknowledgments

We would like to thank Darin Ragozzine, Aviv Ofir, Andrew Youdin, and
Yoram Lithwick for useful discussions. We also thank the anonymous
referee for many insightful comments that helped improve the manuscript. AJ
was partially funded by the European Research Council under ERC Starting Grant
agreement 278675--PEBBLE2PLANET and by the Swedish Research Council (grant
2010--3710). RPC was funded by a Marie-Curie Intra-European Fellowship (grant
252431) under the European Commission’s FP7 framework. MBD was supported by the
Swedish Research Council (grants 2008–-4089 and 2011–-3991). Computer
simulations were performed using the Platon cluster at Lunarc Center for
Scientific and Technical Computing at Lund University. Some simulation hardware
was purchased with grants from the Royal Physiographic Society of Lund.

\end{document}